\documentclass[11pt,aps,prd,preprint,nofootinbib,longbibliography]{revtex4-1}
\usepackage{graphicx}
\usepackage{dcolumn}
\usepackage{bm}
\usepackage{amssymb}
\usepackage{amsmath}
\usepackage{epsfig}    
\usepackage{xcolor}
\usepackage{slashed}
\usepackage{hyperref}
\usepackage{orcidlink}
\usepackage{booktabs}
\allowdisplaybreaks
\newcommand{\be}{\begin{equation}}
\newcommand{\ee}{\end{equation}}
\newcommand{\bes}{\begin{equation*}}
\newcommand{\ees}{\end{equation*}}
\newcommand{\bea}{\begin{eqnarray}}
\newcommand{\eea}{\end{eqnarray}}
\newcommand{\beas}{\begin{eqnarray*}}
\newcommand{\eeas}{\end{eqnarray*}}

\begin{document} 

\title{Phenomenology of Spontaneous CP-violation in 2HDMs}

\author{Tanmoy Mondal \orcidlink{0000-0002-6441-7687}}
\email{tanmoy.mondal@pilani.bits-pilani.ac.in}
\affiliation{Department of Physics, Birla Institute of Technology and Science, Pilani, 333031, Rajasthan, India}

\author{Kei Yagyu \orcidlink{0000-0001-6382-6043}}
\email{yagyu@rs.tus.ac.jp}
\affiliation{Department of Physics, Tokyo University of Science,
1-3, Kagurazaka, Shinjuku-ku, Tokyo 162-8601, Japan}

\begin{abstract}

Spontaneous CP violation (SCPV) provides an attractive possibility for explaining the origin of CP violation and the observed CP-violating phenomena in the quark sector.
We investigate SCPV in the two-Higgs-doublet model as one of its simplest realizations, in which the additional Higgs bosons are necessarily non-decoupling and flavor misalignment in the Yukawa sector is required.
The non-decoupling Higgs sector predicts sizable deviations from the Standard Model predictions even in the Higgs alignment limit: the decay rates of the discovered Higgs boson into $\gamma\gamma$ and $Z\gamma$ can deviate by at least about $-10\%$ and $-4\%$, respectively, while the trilinear Higgs self-coupling can receive ${\cal O}(100)\%$ corrections.
We consider a scenario in which the down-type quark and charged-lepton Yukawa matrices are flavor aligned, while flavor-violating interactions are allowed in the up-type quark sector to reproduce the Kobayashi--Maskawa phase.
We explore the viable parameter space by imposing theoretical constraints together with experimental constraints from electroweak precision observables, flavor measurements, electric dipole moments, and LHC data.
We find that a sizable flavor-violating top--charm Yukawa coupling remains allowed, giving rise to distinctive LHC signatures, including same-sign top-quark and three-top production as well as scalar cascade decays.
We also find that the deviation in the trilinear Higgs self-coupling is correlated with the decay patterns of the additional neutral Higgs bosons, providing a connection between precision Higgs measurements and direct searches for the additional Higgs bosons.
Our results demonstrate the importance of combining searches for additional Higgs bosons with precision Higgs measurements, complemented by flavor and electric-dipole-moment experiments, to probe SCPV.

\end{abstract}
\maketitle

\section{Introduction}

CP violation (CPV) was first established through the discovery of the $K_L\to\pi\pi$ decay, in addition to the dominant $\pi\pi\pi$ mode, by Cronin and Fitch in 1964~\cite{Christenson:1964fg}. Since then, various CP-violating phenomena, including direct CPV, have been observed in the $B$-meson system at the BaBar~\cite{BaBar:2001pki} and Belle~\cite{Belle:2001zzw,Belle:2004hwe} experiments. More recently, the angle $\gamma$ of the Cabibbo--Kobayashi--Maskawa (CKM) unitarity triangle has been precisely measured by the LHCb experiment~\cite{LHCb:2013prk,LHCb:2016mag}.
All these observations of CPV are successfully accommodated by the Kobayashi--Maskawa (KM) phase in the CKM matrix within the three-generation Standard Model (SM)~\cite{Kobayashi:1973fv}.

Despite this remarkable success, the origin of CPV remains unknown. In particular, CPV can arise either explicitly or spontaneously. In the SM, only the former possibility is realized, since the Higgs potential is CP invariant and, apart from the strong CP phase, the only source of CPV originates from the complex phase in the quark Yukawa sector.
On the other hand, both explicit CP violation (ECPV) and spontaneous CP violation (SCPV) can generally be realized in models with extended Higgs sectors because of their richer vacuum structure compared with that of the SM. Extended Higgs sectors are also frequently introduced in new-physics scenarios addressing phenomena unexplained within the SM, such as tiny neutrino masses, dark matter, and the baryon asymmetry of the Universe.

Scenarios with SCPV are particularly well motivated for several reasons.
First, SCPV provides a dynamical origin of CPV through the vacuum structure, in close analogy with electroweak symmetry breaking~\cite{Lee:1973iz}.
Second, SCPV can provide an alternative solution to the strong CP problem~\cite{Nelson:1983zb,Barr:1984qx} to the Peccei--Quinn mechanism~\cite{Peccei:1977hh}.
Third, SCPV naturally leads to the non-decoupling nature of additional Higgs bosons~\cite{Mondal:2025iie}, which can play an important role in realizing a strongly first-order electroweak phase transition, as required for successful electroweak baryogenesis (EWBG)~\cite{Rubakov:1996vz,Cohen:1993nk}.
Moreover, SCPV can provide additional sources of CPV beyond the KM phase, which may also play an important role in EWBG.
Clarifying the origin of CPV is therefore an important avenue for probing physics beyond the SM.

SCPV was first proposed by T.~D.~Lee in the framework of the two-Higgs-doublet model (2HDM)~\cite{Lee:1973iz}, in which the vacuum expectation values (VEVs) of the Higgs doublet fields can develop a nontrivial relative complex phase.
The vacuum structure of the 2HDM has since been extensively studied; see, e.g., Refs.~\cite{Sher:1988mj,Branco:2011iw} for reviews.
In particular, Refs.~\cite{Barroso:2005sm,Barroso:2007rr,Ivanov:2007de} have shown that a CP-violating vacuum, if it exists, is necessarily the global minimum of the potential.
In Ref.~\cite{Gunion:2005ja}, conditions for SCPV were formulated in a basis-independent manner.
The phenomenology of the 2HDM with SCPV has also been studied in Refs.~\cite{Nebot:2018nqn,Nebot:2019qvr,Grzadkowski:2016szj}, with particular emphasis on scenarios with a $Z_2$ symmetry.
It was pointed out in Ref.~\cite{Nierste:2019fbx}, however, that reproducing the KM phase in the 2HDM with SCPV is incompatible with flavor alignment of the Yukawa matrices, owing to the reality of the Yukawa couplings before electroweak symmetry breaking.
Flavor misalignment is therefore an unavoidable feature of such a scenario.
In our previous work~\cite{Mondal:2025iie}, we showed that, in a viable SCPV 2HDM, the properties of the discovered Higgs boson ($h$) can exhibit sizable deviations from the SM predictions as a consequence of the non-decoupling nature of the additional Higgs bosons.
In particular, the decay rate of $h\to\gamma\gamma$ and the trilinear Higgs self-coupling $hhh$ can deviate from their SM predictions by about $-10\%$ and ${\cal O}(100)\%$, respectively.

In this paper, we investigate the collider phenomenology of the SCPV 2HDM, focusing not only on the properties of the discovered Higgs boson but also on direct signatures of the additional Higgs bosons.
In particular, we study their flavor-violating decays, which arise from the constrained structure of the Yukawa sector inherent to SCPV.
By imposing theoretical constraints from vacuum stability and perturbative unitarity, together with experimental constraints from electroweak precision observables, flavor measurements, electric dipole moments (EDMs), Higgs measurements, and direct searches at the LHC, we clarify that an off-diagonal coupling in the up-type quark Yukawa sector, particularly the top--charm flavor-violating interaction, can be sizable and give rise to distinctive signatures at the LHC.
We systematically investigate the possible production and decay channels of the additional Higgs bosons and evaluate the corresponding cross sections for representative benchmark points satisfying all relevant constraints.
We demonstrate that searches for the additional Higgs bosons at the High-Luminosity LHC (HL-LHC) can probe a significant portion of the viable parameter space of the SCPV 2HDM.
Furthermore, we identify a characteristic correlation between deviations in the trilinear Higgs self-coupling from the SM prediction and the decay patterns of the additional Higgs bosons.

This paper is organized as follows.
In Sec.~\ref{sec:model}, we briefly review SCPV in the 2HDM. We first discuss the vacuum structure and then present the Yukawa interactions, which inevitably contain flavor-violating interactions in the SCPV scenario.
In Sec.~\ref{sec:constraints}, we discuss theoretical and experimental constraints on the model parameter space.
In Sec.~\ref{sec:phenomenology}, we investigate the collider phenomenology of the additional Higgs bosons at the LHC.
Finally, conclusions are given in Sec.~\ref{sec:discussions}.

\section{Model \label{sec:model}}

\subsection{Higgs potential and spontaneous CP-violation}

We briefly review the 2HDM with SCPV. The Higgs sector consists of two $SU(2)_L$ scalar doublets, $\Phi_1$ and $\Phi_2$, both with hypercharge $Y=1/2$. The most general renormalizable Higgs potential is given by
\begin{align}
V&=m_1^2|\Phi_1|^2 + m_2^2|\Phi_2|^2-(m_3^2\Phi_1^\dagger \Phi_2 +\text{h.c.})+\frac{\lambda_1}{2}|\Phi_1|^4+\frac{\lambda_2}{2}|\Phi_2|^4+\lambda_3|\Phi_1|^2|\Phi_2|^2+\lambda_4|\Phi_1^\dagger\Phi_2|^2\notag\\
& 
+ \left[\frac{\lambda_5}{2}(\Phi_1^\dagger\Phi_2)+\lambda_6|\Phi_1|^2+\lambda_7|\Phi_2|^2\right](\Phi_1^\dagger\Phi_2)+\text{h.c.},  \label{pot_thdm1}
\end{align}
where $m_3^2$ and $\lambda_{5,6,7}$ are generally complex parameters.
Since we focus on SCPV, we work in a basis where all the parameters in the Higgs potential are real.\footnote{A Higgs potential with complex parameters can be transformed into a real form by an appropriate $U(2)$ basis transformation, provided that the potential is explicitly CP conserving. See Ref.~\cite{Gunion:2005ja} for basis-independent conditions for explicit and spontaneous CP violation in the 2HDM.}
Assuming an electrically neutral vacuum that preserves $U(1)_{\rm em}$, the VEVs of the two doublets can be parameterized as~\cite{Sher:1988mj}
\begin{align}
\langle\Phi_1\rangle = 
\begin{pmatrix}
0\\
\frac{v_1}{\sqrt{2}}
\end{pmatrix},\quad 
\langle\Phi_2\rangle = 
\begin{pmatrix}
0\\
\frac{v_2}{\sqrt{2}}e^{i\xi}
\end{pmatrix}, \quad 
v_1,v_2,\xi\in \mathbb{R}. \label{eq:vev}
\end{align}
A nontrivial relative phase, $\xi\neq n\pi$ with $n\in\mathbb{Z}$, allows for SCPV~\cite{Gunion:2005ja}. The VEVs satisfy $v^2\equiv v_1^2+v_2^2=(\sqrt{2}G_F)^{-1}$, where $G_F$ is the Fermi constant.
The stationary conditions evaluated at this vacuum are given by
\begin{align}
\begin{split}
\frac{\partial V_0}{\partial v_1} &= 
v_1T_1
-v_2c_\xi T_3
=0,\\
\frac{\partial V_0}{\partial v_2}&=
v_2T_2 - v_1c_\xi T_3 = 0,\\
\frac{\partial V_0}{\partial \xi} &= 
v_1v_2s_\xi T_3 = 0, 
\end{split}
\label{eq:sc}
\end{align}
where we use the shorthand notation $c_X\equiv\cos X$ and $s_X\equiv\sin X$. Here, $V_0$ denotes the Higgs potential evaluated at the vacuum defined in Eq.~(\ref{eq:vev}), and $T_{1,2,3}$ are defined as
\begin{align}
T_1 &= m_1^2 + \frac{v_1^2}{2}\lambda_1 + \frac{v_2^2}{2}(\lambda_3+\lambda_4 -\lambda_5) + v_1v_2c_\xi \lambda_6, \\
T_2 & = m_2^2 + \frac{v_2^2}{2}\lambda_2 + \frac{v_1^2}{2}(\lambda_3+\lambda_4 -\lambda_5) + v_1v_2c_\xi \lambda_7, \\
T_3 &= m_3^2 -v_1v_2c_\xi \lambda_5-\frac{v_1^2}{2}\lambda_6 - \frac{v_2^2}{2}\lambda_7. 
\end{align}
For $v_1v_2s_\xi\neq0$, the stationary conditions in Eq.~(\ref{eq:sc}) reduce to $T_1=T_2=T_3=0$. These conditions can be solved for the three quadratic mass parameters $m_i^2$ ($i=1,2,3$), yielding
\begin{align}
\begin{pmatrix}
m_1^2\\
m_2^2\\
m_3^2
\end{pmatrix}
=
\begin{pmatrix}
-\frac{\lambda_1}{2} & -\frac{1}{2}(\lambda_3+\lambda_4-\lambda_5) & -\lambda_6\\
-\frac{1}{2}(\lambda_3+\lambda_4-\lambda_5) & -\frac{\lambda_2}{2} & -\lambda_7 \\
\frac{\lambda_6}{2} & \frac{\lambda_7}{2} & \lambda_5
\end{pmatrix}
\begin{pmatrix}
v_1^2\\
v_2^2\\
v_1v_2c_\xi
\end{pmatrix}. \label{eq:sc2}
\end{align}
Eq.~(\ref{eq:sc2}) shows that the quantities $v_1^2$, $v_2^2$, and $v_1v_2c_\xi$ are uniquely determined by the parameters of the Higgs potential, provided that the $3\times3$ coefficient matrix is invertible. Consequently, the CP-violating stationary point is unique up to symmetry transformations, as discussed in Ref.~\cite{Barroso:2005sm}.
Furthermore, it has been shown that a CP-violating minimum cannot coexist with a deeper CP-conserving or charge-breaking minimum~\cite{Ferreira:2004yd,Barroso:2005sm,Barroso:2007rr,Ivanov:2007de}. Thus, if the CP-violating stationary point is a local minimum, it is also the global minimum, up to possible symmetry-related degeneracies.
This feature contrasts with the CP-conserving case, where $s_\xi=0$. In this case, the third stationary condition in Eq.~(\ref{eq:sc}) is automatically satisfied, and multiple CP-conserving stationary points may exist.
Another important consequence of Eq.~(\ref{eq:sc2}) is the absence of a decoupling limit in the SCPV scenario. All three quadratic mass parameters $m_i^2$ are determined by the VEVs and the quartic couplings $\lambda_j$ ($j=1,\ldots,7$). Since the quartic couplings are bounded by perturbative unitarity and perturbativity~\cite{Nierste:2019fbx,Nebot:2019qvr}, the quadratic mass parameters cannot be made arbitrarily large compared with the electroweak scale. As a result, the additional Higgs bosons cannot be arbitrarily heavy, and their masses are typically constrained to lie below the TeV scale.
This non-decoupling property can be explicitly understood from the Higgs boson mass formulas presented below.

We expand the scalar doublets around the CP-violating vacuum, and they can be expressed 
in the Higgs basis~\cite{Davidson:2005cw} as
\begin{align}
\left(\begin{array}{c}
\Phi_1\\
\Phi_2
\end{array}\right)=
\begin{pmatrix}
1 & 0\\
0 & e^{i\xi}
\end{pmatrix}
\left(\begin{array}{cc}
\cos\beta & -\sin\beta\\
\sin\beta & \cos\beta
\end{array}\right)
\left(\begin{array}{c}
\Phi\\
\Phi'
\end{array}\right), \label{eq:hb2}
\end{align}
where $\tan\beta=v_2/v_1$ and
\begin{align}
\Phi=\left[
\begin{array}{c}
G^+\\
\frac{1}{\sqrt{2}}(h_1'+v+ iG^0)
\end{array}\right],\quad
\Phi'=\left[
\begin{array}{c}
H^+\\
\frac{1}{\sqrt{2}}(h_2'+ih_3')
\end{array}\right]. \label{Higgs-basis}
\end{align}
Here, $G^\pm$ and $G^0$ are the Nambu–Goldstone (NG) bosons, $H^\pm$ denotes the physical charged Higgs bosons, and $h_1'$, $h_2'$, and $h_3'$ represent the neutral scalar fields. In general, these three neutral fields mix with one another because of the CP-violating vacuum.

The squared mass of the charged Higgs bosons is given by
\begin{align}
m_{H^\pm}^2 & = \frac{v^2}{2}(\lambda_5 - \lambda_4). \label{eq:mch}
\end{align}
The charged Higgs bosons become massless in the limit $\lambda_4=\lambda_5$. This behavior can be understood in terms of an enhanced global symmetry of the Higgs potential. In this limit, the potential possesses a global $SU(2)_L\times SU(2)_R$ symmetry, under which the bidoublet fields $M_i\equiv(i\tau_2\Phi_i^*,\Phi_i)$ transform as $M_i\to U_LM_iU_R^\dagger$~\cite{Pomarol:1993mu,Aiko:2020atr}.
The vacuum configurations are given by $\langle M_1\rangle=v_1 I_{2\times2}/\sqrt{2}$ and $\langle M_2\rangle=v_2\,\mathrm{diag}(e^{-i\xi},e^{i\xi})/\sqrt{2}$. For a nontrivial relative phase $\xi$, the global symmetry is spontaneously broken down to $U(1)$, giving rise to two additional massless NG bosons, which are identified with $H^\pm$.

The squared-mass matrix for the neutral Higgs bosons, $\mathcal{M}$, in the basis $(h_1',h_2',h_3')$ is given by
\begin{align}
\begin{split}   
\frac{{\cal M}_{11}}{v^2} & = \lambda_1 c_\beta^4 + \lambda_2s_\beta^4
+2(\lambda_3 +\lambda_4 + \lambda_5 c_{2\xi}  )c_\beta^2 s_\beta^2
+2(\lambda_6c_\beta^2 + \lambda_7 s_\beta^2)s_{2\beta}c_\xi, \\
\frac{{\cal M}_{22}}{v^2} & = (\lambda_1 + \lambda_2 -2 \lambda_3  -2 \lambda_4 )c_\beta^2s_\beta^2  + \frac{\lambda_5}{2}(1 + c_{2\beta}^2c_{2\xi}) - (\lambda_6 - \lambda_7)s_{2\beta}c_{2\beta}c_\xi, \\
\frac{{\cal M}_{33}}{v^2} & = \lambda_5 s_\xi^2, \\
\frac{{\cal M}_{12}}{v^2} & = (\lambda_2s_\beta^2 - \lambda_1c_\beta^2)c_\beta s_\beta + \frac{1}{4}(\lambda_3 + \lambda_4 + \lambda_5 c_{2\xi})s_{4\beta}
 + [\lambda_6(2c_{2\beta} - 1)c_\beta^2 + \lambda_7(2c_{2\beta} + 1)s_\beta^2] c_\xi, \\
\frac{{\cal M}_{13}}{v^2} & = -\frac{\lambda_5}{2}s_{2\beta}s_{2\xi}-(\lambda_6c_\beta^2 + \lambda_7s_\beta^2)s_\xi, \\
\frac{{\cal M}_{23}}{v^2} & =-\frac{\lambda_5}{2} c_{2\beta}s_{2\xi} + \frac{1}{2}(\lambda_6 - \lambda_7)s_{2\beta}s_\xi. 
\end{split}\label{eq:mat-hb}
\end{align}
The neutral Higgs mass eigenstates are defined by
\begin{align}
\begin{pmatrix}
h_1' \\
h_2' \\
h_3' 
\end{pmatrix} 
= 
R
\begin{pmatrix}
h \\
H_2 \\
H_3 
\end{pmatrix}, \label{eq:rot}
\end{align}
where $R$ is a $3\times3$ orthogonal matrix that diagonalizes $\mathcal{M}$:
\begin{align}
R^T {\cal M} R = \text{diag}(m_{h}^2,m_{H_2}^2,m_{H_3}^2).  
\end{align}
We identify $h$ with the observed Higgs boson with a mass of $125$ GeV, i.e., $m_h=125$ GeV.
Taking the trace of the squared-mass matrix, we obtain the sum of the squared masses of the three neutral Higgs bosons:
\begin{align}
\text{tr}[{\cal M} ]= v^2[\lambda_1 c_\beta^2 + \lambda_2 s_\beta^2 + \lambda_5 + (\lambda_6 + \lambda_7) s_{2\beta}c_\xi]. 
\end{align}
As discussed above, the squared masses of the Higgs bosons are determined entirely by the electroweak VEV and the quartic couplings. Therefore, the SCPV scenario does not admit a decoupling limit in which the additional Higgs bosons become arbitrarily heavy while the quartic couplings remain perturbative.
Furthermore, the determinant of the squared-mass matrix satisfies
\begin{align}
\text{det}[{\cal M}] \propto  s_\beta^2 c_\beta^2s_\xi^2 = \frac{s_\xi^2}{(t_\beta + 1/t_\beta)^2}. 
\end{align}
For fixed finite quartic couplings, this relation implies that at least one neutral Higgs boson becomes massless in the limits $s_\xi\to0$, $t_\beta\to0$, or $t_\beta\to\infty$. Consequently, requiring all three neutral Higgs bosons to have nonzero masses imposes restrictions on the allowed ranges of $\xi$ and $\tan\beta$.

The Higgs alignment limit is realized when the neutral scalar field $h_1'$, which corresponds to the fluctuation along the direction of the VEV, does not mix with the other neutral fields. This condition is expressed as
\begin{align}
{\cal M}_{12} =  {\cal M}_{13} = 0. 
\end{align}
These conditions can be solved for $\lambda_6$ and $\lambda_7$ as
\begin{align}
\lambda_6 &= \left[-\lambda_1c_{\beta} + \lambda_2t_\beta s_\beta + (\lambda_3 + \lambda_4)\frac{c_{2\beta}}{c_\beta} +\frac{\lambda_5}{c_\beta}(1 - 2c_{\beta}^2(2 + c_{2\xi}))\right]\frac{s_\beta}{2c_\xi}, \\
\lambda_7 &= \left[ \lambda_1 \frac{c_\beta}{t_\beta} -\lambda_2 s_\beta -(\lambda_3 + \lambda_4) \frac{c_{2\beta}}{s_\beta} + \frac{\lambda_5}{s_\beta}(1- 2s^2_\beta(2+c_{2\xi}))\right]\frac{c_\beta}{2c_\xi}. 
\end{align}
In the alignment limit, the orthogonal matrix $R$ reduces to
\begin{align}
R \to
\begin{pmatrix}
1 & 0 & 0 \\
0 & c_\alpha & -s_\alpha \\
0 & s_\alpha & c_\alpha
\end{pmatrix}, 
\end{align}
where $\alpha$ describes the mixing between the two additional neutral Higgs bosons. In this limit, the couplings of $h$ to the SM gauge bosons and fermions coincide with their SM values at tree level.
The remaining quartic couplings, $\lambda_1,\ldots,\lambda_5$, can be expressed in terms of the physical Higgs boson masses and the mixing angles as follows:
\begin{align}
\lambda_1 v^2 & = m_h^2 + m_{H_2}^2 \left(\frac{t_\beta s_{\alpha - \xi}}{s_\xi}\right)^2 + m_{H_3}^2 \left(\frac{t_\beta c_{\alpha - \xi}}{s_\xi}\right)^2, \\
\lambda_2v^2 & = m_h^2 + m_{H_2}^2 \left(\frac{s_{\alpha + \xi}}{t_\beta s_\xi}\right)^2 + m_{H_3}^2 \left(\frac{ c_{\alpha + \xi}}{t_\beta s_\xi}\right)^2, \\
\lambda_3v^2 & = m_h^2 + 2m_{H^\pm}^2 + \frac{1}{2s_\xi^2}[m_{H_2}^2(c_{2\xi}-c_{2\alpha}) + m_{H_3}^2(c_{2\xi} + c_{2\alpha})], \\
\lambda_4v^2 & = -2m_{H^\pm}^2 + \frac{1}{s_\xi^2}(m_{H_2}^2s_\alpha^2+ m_{H_3}^2c_{\alpha}^2), \\
\lambda_5v^2 & =  \frac{1}{s_\xi^2}(m_{H_2}^2s_\alpha^2+ m_{H_3}^2c_{\alpha}^2). 
\end{align}
These relations explicitly demonstrate that the quartic couplings grow with the squared masses of the additional Higgs bosons. Consequently, perturbative unitarity and vacuum stability constraints can be translated into restrictions on the allowed Higgs boson masses and mixing parameters. The relevant theoretical constraints are discussed in detail in Sec.~\ref{sec:constraints}.

In the Higgs alignment limit, the relevant scalar trilinear couplings take particularly simple forms:
\begin{align}
\lambda_{H^+H^-h} & = \frac{1}{v}(2m_{H^\pm}^2 + m_h^2),
\label{eq:hphmh}
\\
\lambda_{H_iH_ih} & = \frac{1}{2v}(2m_{H_i}^2 + m_h^2),  ~~~(i = 2,3), \\
\lambda_{H_ihh} & = 0,  ~~~(i = 2,3), \\
\lambda_{hhh} & = \frac{m_h^2}{2v}, \\
\lambda_{H^+H^-H_2}& = \frac{m_{H_2}^2}{v s_\beta c_\beta}(c_\alpha c_{2\beta}+s_\alpha \cot\xi), \\
\lambda_{H^+H^-H_3}& = \frac{m_{H_3}^2}{v s_\beta c_\beta}(c_\alpha \cot\xi -s_\alpha c_{2\beta}), \\
\lambda_{H_3H_3H_2}& =\frac{\lambda_{H^+H^-H_3}}{2}, \\
\lambda_{H^\pm G^\mp h} & = 0\\
\lambda_{H^\pm G^\mp H_2} & = \frac{e^{-i\alpha}}{v}(m_{H_2}^2 - m_{H^\pm}^2),\\
\lambda_{H^\pm G^\mp H_3} & = -i\frac{e^{-i\alpha}}{v}(m_{H_3}^2 - m_{H^\pm}^2).
\end{align}
Here, $\lambda_{ijk}$ denotes the coefficient of the corresponding trilinear scalar interaction in the Higgs potential, e.g., $V \supset \lambda_{hhh}hhh$. 
Among these couplings, the Higgs trilinear self-coupling $\lambda_{hhh}$ is of particular importance, as it provides a direct probe of the Higgs potential and is expected to be measured at the HL-LHC and future lepton colliders. Although its tree-level value coincides with the SM prediction in the alignment limit, sizable deviations can arise from radiative corrections involving the additional Higgs bosons.
Using the effective potential method~\cite{Coleman:1973jx}, the one-loop contribution from the additional Higgs bosons to the trilinear Higgs self-coupling can be expressed as~\cite{Mondal:2025iie}
\begin{align}
\lambda_{hhh}^{\rm 1-loop}-\lambda_{hhh}^{\rm 1-loop}\Big|_{\rm SM} &= \frac{1}{16\pi^2}\left(\frac{\lambda_{H^+H^-h}^3}{m_{H^\pm}^2} + \frac{4\lambda_{H_2H_2h}^3}{m_{H_2}^2} + \frac{4\lambda_{H_3H_3h}^3}{m_{H_3}^2}\right)\notag\\
& = \frac{1}{32\pi^2}\sum_{\phi=H^\pm,H_2,H_3}n_\phi\frac{(2m_\phi^2 + m_h^2)^3}{v^3m_\phi^2}, \label{eq:hhh}
\end{align}
where $n_{H^\pm}=2$ and $n_{H_2}=n_{H_3}=1$ count the real scalar degrees of freedom.
The correction exhibits a characteristic non-decoupling behavior. In particular, for $m_\phi^2\gg m_h^2$, the contribution of each additional Higgs boson scales as $m_\phi^4/v^3$. Thus, sizable deviations from the SM prediction can arise even in the exact Higgs alignment limit, providing a complementary probe of SCPV through precision measurements of the Higgs self-coupling.

\subsection{Yukawa interaction}

The most general Yukawa interactions in the SCPV 2HDM are given by
\begin{align}
&\mathcal{L}_Y=
-\sum_{\alpha = 1,2}\Bigg[Y_d^\alpha \left(\overline{\tilde{Q}_L}\Phi_\alpha \tilde{d}_R^{}\right)
+Y_u^\alpha\left(\overline{\tilde{Q}_L} \Phi^c_\alpha \tilde{u}_R^{}\right)
+Y_e^\alpha\left(\overline{\tilde{L}_L} \Phi_\alpha \tilde{e}_R^{}\right)
+\text{h.c.}\Bigg],\quad  \label{Eq:Yukawa0}
\end{align}
where $\tilde{Q}_L$ and $\tilde{L}_L$ denote the left-handed quark and lepton doublets, respectively, while $\tilde{d}_R$, $\tilde{u}_R$, and $\tilde{e}_R$ denote the corresponding right-handed fermion singlets. The fermion fields with tildes are defined in the weak eigenbasis. The charge-conjugated Higgs doublets are defined as $\Phi_\alpha^c\equiv i\tau_2\Phi_\alpha^*$.
The Yukawa matrices $Y_f^\alpha$ ($f=d,u,e$) are general $3\times3$ matrices. To realize SCPV without ECPV   in the Yukawa sector, we impose the reality condition
\begin{align}
Y_f^\alpha \in \mathbb{R}. \label{eq:reality}
\end{align}
After electroweak symmetry breaking, the fermion mass matrices are generated by linear combinations of the two Yukawa matrices:
\begin{align}
M_f&=\frac{v}{\sqrt{2}}(Y_f^1c_\beta + Y_f^2e^{-2iI_f \xi}s_\beta),  \label{Eq:Mf} 
\end{align}
where $I_f=1/2$ for $f=u$ and $I_f=-1/2$ for $f=d,e$.
Although $Y_f^1$ and $Y_f^2$ are real, the resulting mass matrices $M_f$ are generally complex due to the relative phase $\xi$ of the Higgs VEVs. For $v_1v_2s_\xi\neq0$, the two independent real Yukawa matrices can reproduce an arbitrary complex mass matrix. Consequently, the observed fermion masses and the KM phase can be accommodated, with the latter originating entirely from the spontaneous breaking of CP symmetry.
In the Higgs basis and the fermion mass eigenbasis, the Yukawa interactions can be rewritten as
\begin{align}
\mathcal{L}_Y &=
-\overline{Q_L^d}\left(\frac{\sqrt{2}}{v}M_d^{\rm diag} \Phi + \rho_d\Phi'\right)d_R^{}
-\overline{Q_L^u}\left(\frac{\sqrt{2}}{v}M_u^{\rm diag} \Phi^c + \rho_u \Phi^{\prime c}\right)u_R^{} \notag\\
&-\overline{L_L}\left(\frac{\sqrt{2}}{v}M_e^{\rm diag} \Phi + \rho_e\Phi'\right) e_R^{}
+\text{h.c.},  \label{BottomUp:2HDM:Yukawa2}
\end{align}
where the fermion mass eigenstates are obtained through the biunitary transformations $\tilde f_L=V_f f_L$ and $\tilde f_R=U_f f_R$, with $V_f$ and $U_f$ being unitary matrices. The left-handed quark doublets are defined as
\begin{align}
Q_L^d = 
\begin{pmatrix}
V_{\rm CKM}^\dagger u_L \\
d_L
\end{pmatrix},\quad
Q_L^u = 
\begin{pmatrix}
u_L \\
V_{\rm CKM} d_L
\end{pmatrix}, 
\end{align}
where $V_{\rm CKM}\equiv V_u^\dagger V_d$ is the CKM matrix.
In Eq.~(\ref{BottomUp:2HDM:Yukawa2}), $M_f^{\rm diag}$ denote the diagonal fermion mass matrices with real and positive entries, while $\rho_f$ are the Yukawa coupling matrices associated with the second Higgs doublet in the Higgs basis.
In a general 2HDM, $\rho_f$ are arbitrary complex $3\times3$ matrices. In contrast, the reality of $Y_f^{1,2}$ in the SCPV scenario imposes the following relations:
\begin{align}
\rho_f  &= \frac{1}{\sqrt{2}c_\beta s_\beta v}\left[\left(c_{2\beta} +2I_f\frac{i}{t_\xi}\right) M_f^{\rm diag} 
- \left(1 + 2I_f\frac{i}{t_\xi}\right)(V_f^\dagger V_f^* M_f^{\rm diag} U_f^TU_f)\right]. \label{eq:yf}
\end{align}
The off-diagonal elements of $\rho_f$ generally induce flavor-changing neutral currents (FCNCs) mediated by the neutral Higgs bosons at tree level.

Tree-level FCNCs can be avoided by imposing Yukawa alignment~\cite{Pich:2009sp}, in which the two Yukawa matrices for each fermion species are proportional to each other:
\begin{align}
Y_f^2 = \tilde{\zeta}_f Y_f^1,\quad \tilde\zeta \in \mathbb{R}.
\end{align}
Under this assumption, the fermion mass matrices take the form
\begin{align}
M_f = \frac{v}{\sqrt{2}}|z_f|e^{i\arg z_f}\,Y_f^1,~~z_f = c_\beta + e^{-2iI_f \xi}\tilde{\zeta}_fs_\beta,~~\arg z_f = -2I_f\arctan\left(\frac{t_\beta s_\xi \tilde{\zeta}_f}{1 + t_\beta c_\xi \tilde{\zeta}_f}\right). 
\end{align}
The complex phase then appears only as an overall factor, $e^{i\arg z_f}$, multiplying the real Yukawa matrix $Y_f^1$. This phase can be removed by rephasing the fermion fields, so that the KM phase vanishes if Yukawa alignment is imposed in both the up- and down-type quark sectors~\cite{Nierste:2019fbx}.
To reproduce the observed KM phase while suppressing potentially dangerous FCNCs, we follow the scenario proposed in Ref.~\cite{Mondal:2025iie}, in which Yukawa alignment is imposed only in the down-type quark and charged-lepton sectors, while the up-type quark Yukawa matrices remain general.
In the aligned sectors, the biunitary transformations can be expressed in terms of real orthogonal matrices:
\begin{align}
V_f \to O_f^L,\quad U_f \to O_f^R e^{-i\arg z_f}~~(f=d,e), 
\end{align}
where $O_f^L$ and $O_f^R$ are orthogonal matrices.
Consequently, the Yukawa coupling matrices $\rho_d$ and $\rho_e$ become flavor diagonal:
\begin{align}
\rho_f  &= \frac{\sqrt{2}\zeta_f }{v}M_f^{\rm diag} ~~(f=d,e), \label{eq:rho_dee}
\end{align}
where the complex alignment parameters $\zeta_f$ are given by
\begin{align}\label{eq:zeta-d-e}
\zeta_f  
=   \frac{-t_\beta +  \tilde{\zeta}_{f}\,e^{-2iI_f\xi}}{1+  t_\beta \,\tilde{\zeta}_{f}\,e^{-2iI_f\xi}} ~~(f=d,e).
\end{align}
In the limiting cases $\tilde{\zeta}_f\to0$ and $\tilde{\zeta}_f\to\infty$, these parameters reduce to
\begin{align}
\zeta_f & \to 
\begin{cases}
-\tan\beta~~\text{for}~~\tilde{\zeta}_f \to 0~(\arg z_f\to 0)\\
\cot\beta~~\text{for}~~\tilde{\zeta}_f \to \infty ~(\arg z_f\to -2I_f\xi)
\end{cases}~~(f=d,e), 
\end{align}
corresponding to $Y_f^2=0$ and $Y_f^1=0$, respectively, as realized by appropriate $Z_2$-symmetric Yukawa structures.
Using Eq.~(\ref{eq:yf}) together with $V_u=O_d^L V_{\rm CKM}^\dagger$, we obtain the following expression for the up-type Yukawa coupling matrix:
\begin{align}
\rho_u  &= \frac{\sqrt{2}}{s_{2\beta} v}\left[\left(c_{2\beta} +\frac{i}{t_\xi}\right) M_u^{\rm diag}  
- \left(1 + \frac{i}{t_\xi}\right)(V_{\rm CKM} V_{\rm CKM}^T M_u^{\rm diag} U_u^TU_u)\right],  \label{eq:rho_u}
\end{align}
The unitary matrix $U_u$ is parameterized as
\be
U_{u} =
\begin{pmatrix}
1 & 0 & 0 \\
0 & \cos\theta_{23} & \sin\theta_{23} e^{-i \delta_{23}} \\
0 & -\sin\theta_{23} e^{i \delta_{23}} & \cos\theta_{23}
\end{pmatrix}
\begin{pmatrix}
\cos\theta_{13} & 0 & \sin\theta_{13} e^{-i \delta_{13}} \\
0 & 1 & 0 \\
-\sin\theta_{13} e^{i \delta_{13}} & 0 & \cos\theta_{13}
\end{pmatrix}
\begin{pmatrix}
\cos\theta_{12} & \sin\theta_{12} e^{-i \delta_{12}} & 0 \\
-\sin\theta_{12} e^{i \delta_{12}} & \cos\theta_{12} & 0 \\
0 & 0 & 1
\end{pmatrix}. 
\ee

We now discuss the flavor structure of $\rho_u$, focusing on the flavor-violating contributions arising from the second term in Eq.~(\ref{eq:rho_u}). For this purpose, we introduce the dimensionless matrix
$\rho_u^{\rm FV} \equiv V_{\rm CKM}V_{\rm CKM}^TM_u^{\rm diag}U_u^TU_u/m_t$ \footnote{The contribution from the second term of Eq.~(\ref{eq:rho_u}) is given by
$-\frac{\sqrt{2}m_t}{s_{2\beta}v}\left(1+\frac{i}{t_\xi}\right)\rho_u^{\rm FV}$.
The magnitude of the overall coefficient is approximately $1/|s_{2\beta}s_\xi|$, which is of order unity unless $|s_{2\beta}|$ or $|s_\xi|$ is small.
}. 
First, we consider the case where $U_u$ is real and orthogonal, corresponding to $\delta_{12}=\delta_{13}=\delta_{23}=0$. In this case, $U_u^TU_u$ reduces to the identity matrix, and the flavor structure of $\rho_u^{\rm FV}$ is determined entirely by the CKM matrix and the up-type quark masses. Its elements are estimated as
\begin{align}
\rho_u^{\rm FV}=
\begin{pmatrix}
{\cal O}(r_u) &{\cal O}( r_c\lambda^5) & {\cal O}(\lambda^3)\\
{\cal O}(r_u\lambda^5) &{\cal O}( r_c) & {\cal O}(\lambda^4)\\
{\cal O}(r_u\lambda^3) &{\cal O}( r_c\lambda^4) & {\cal O}(1)
\end{pmatrix}
=\begin{pmatrix}
{\cal O}(10^{-5}) &{\cal O}(10^{-6}) & {\cal O}(10^{-2})\\
{\cal O}(10^{-9}) &{\cal O}(10^{-2}) & {\cal O}(10^{-3})\\
{\cal O}(10^{-7}) &{\cal O}(10^{-5}) & {\cal O}(1)
\end{pmatrix}, \label{eq:rho_ufv}
\end{align}
where $\lambda\simeq0.225$ is the Cabibbo parameter~\cite{Cabibbo:1963yz}, and $r_u=m_u/m_t$ and $r_c=m_c/m_t$ denote the up- and charm-quark mass ratios, respectively.
As shown in Eq.~(\ref{eq:rho_ufv}), the off-diagonal elements are naturally suppressed by powers of the Cabibbo parameter and/or the small quark mass ratios. In particular, the product $|(\rho_u)_{12}(\rho_u)_{21}|$ is strongly constrained by $D^0$–$\bar D^0$ mixing. For additional neutral Higgs boson masses of order $500$ GeV, this product is constrained to be below approximately $\mathcal{O}(10^{-8})$~\cite{Crivellin:2013wna}. Such a constraint can be readily satisfied in the present scenario with a real orthogonal $U_u$, owing to the strong suppression of the relevant off-diagonal elements.
Since flavor violation involving the second and third generations is generally less constrained than that involving the first and second generations, we next consider a simplified scenario with $\theta_{12}=\theta_{13}=\delta_{12}=\delta_{13}=0$. The matrix $U_u$ then reduces to
\be
U_{u} =
\begin{pmatrix}
1 & 0 & 0 \\
0 & \cos\theta_{23} & \sin\theta_{23}  e^{-i \delta_{23}} \\
0 & -\sin\theta_{23}  e^{i \delta_{23}} &\cos\theta_{23} 
\end{pmatrix}. 
\ee
In this case, the elements of $\rho_u^{\rm FV}$ depend on $\theta_{23}$ and $\delta_{23}$ as follows:
\begin{align}
(\rho_u^{\rm FV})_{i1}
&= [(\rho_u^{\rm FV})_{i1}]_{\delta_{23}=0},\\ 
(\rho_u^{\rm FV})_{i2} &= (\cos^2\theta_{23} +\sin^2\theta_{23}e^{2i\delta_{23}})[(\rho_u^{\rm FV})_{i2}]_{\delta_{23}=0}
-i\sin2\theta_{23} \sin\delta_{23}[(\rho_u^{\rm FV})_{i3}]_{\delta_{23}=0}, \\
(\rho_u^{\rm FV})_{i3} &=(\cos^2\theta_{23}+\sin^2\theta_{23}e^{-2i\delta_{23}})[(\rho_u^{\rm FV})_{i3}]_{\delta_{23}=0}
-i\sin2\theta_{23}\sin\delta_{23}[(\rho_u^{\rm FV})_{i2}]_{\delta_{23}=0}. \label{eq:app}
\end{align}
In the limit $\delta_{23}\to0$, the matrix $\rho_u^{\rm FV}$ reduces to the structure given in Eq.~(\ref{eq:rho_ufv}).
Of particular interest is the element $(\rho_u^{\rm FV})_{32}$, which receives a potentially large contribution from $(\rho_u^{\rm FV})_{33}$ due to the top-quark mass enhancement. Neglecting the contributions suppressed by the lighter quark masses and CKM mixing, we obtain
\begin{align}
(\rho_u^{\rm FV})_{32} &\simeq 
-i\sin2\theta_{23} \sin\delta_{23}. 
\end{align}
Thus, $|(\rho_u^{\rm FV})_{32}|$ can reach $\mathcal{O}(1)$ for $\theta_{23}\simeq\pi/4$ and $\delta_{23}\simeq\pi/2$. This leads to potentially sizable flavor-violating interactions between the top and charm quarks, giving rise to characteristic collider signatures discussed in Sec.~\ref{sec:phenomenology}.

\begin{figure}
\includegraphics[width=7cm]{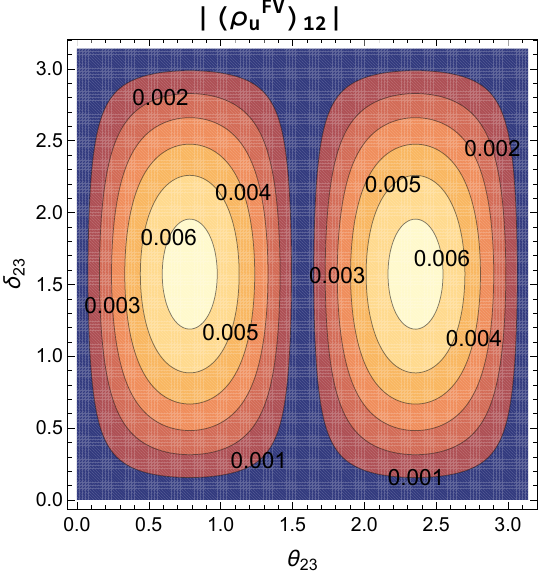}\hspace{5mm}
\includegraphics[width=7cm]{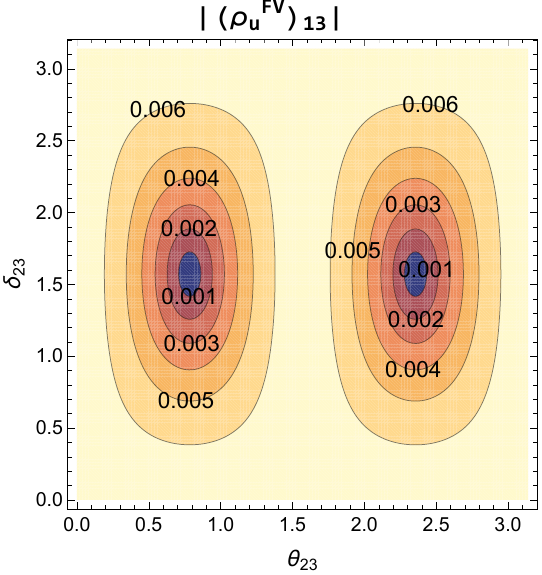}\\
\vspace{5mm}
\includegraphics[width=7cm]{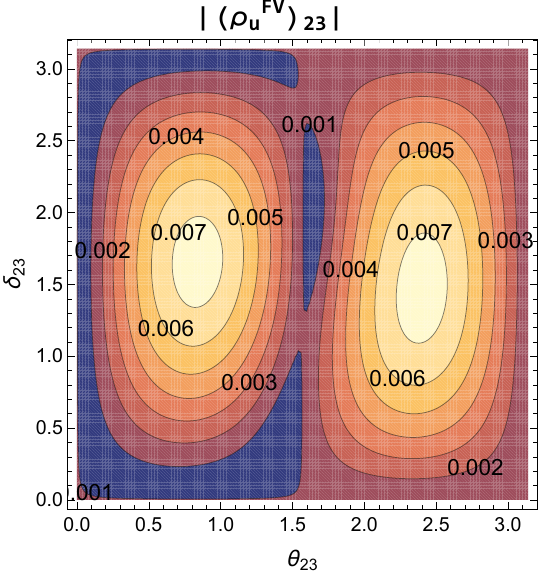}\hspace{5mm}
\includegraphics[width=7cm]{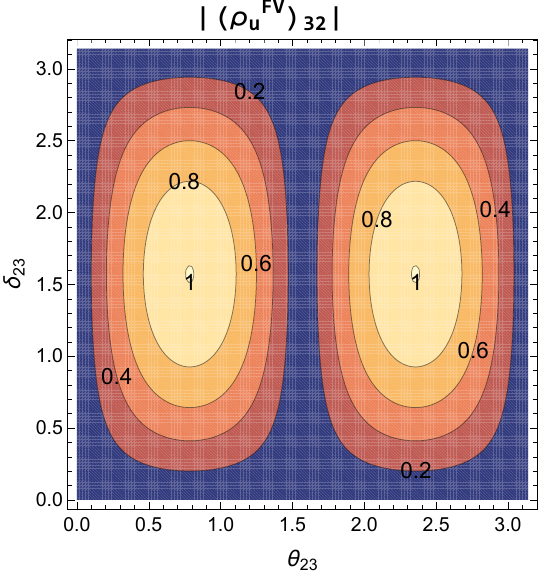}
\caption{Absolute values of the off-diagonal elements of $\rho_u^{\rm FV}$ in the $\theta_{23}$–$\delta_{23}$ plane, with $\theta_{12}=\theta_{13}=\delta_{12}=\delta_{13}=0$. The upper-left, upper-right, lower-left, and lower-right panels show $|(\rho_u^{\rm FV})_{12}|$, $|(\rho_u^{\rm FV})_{13}|$, $|(\rho_u^{\rm FV})_{23}|$, and $|(\rho_u^{\rm FV})_{32}|$, respectively. We use the Wolfenstein parameters $\lambda=0.225$, $A=0.826$, $\rho=0.159$, and $\eta=0.348$, together with the quark masses $m_u=0.0022$ GeV, $m_c=1.27$ GeV, and $m_t=172.6$ GeV.}
\label{fig:rho_u}
\end{figure}

Fig.~\ref{fig:rho_u} shows the absolute values of the off-diagonal elements $|(\rho_u^{\rm FV})_{12}|$ (upper left), $|(\rho_u^{\rm FV})_{13}|$ (upper right), $|(\rho_u^{\rm FV})_{23}|$ (lower left), and $|(\rho_u^{\rm FV})_{32}|$ (lower right) in the $\theta_{23}$–$\delta_{23}$ plane.
As expected from the analytical discussion, $|(\rho_u^{\rm FV})_{32}|$ can reach values of order unity due to the contribution proportional to the top-quark mass. In contrast, the other off-diagonal elements shown in the figure remain typically at or below the $\mathcal{O}(10^{-3})$ level.
We also note that $|(\rho_u^{\rm FV})_{21}|$ and $|(\rho_u^{\rm FV})_{31}|$ are independent of $\theta_{23}$ and $\delta_{23}$, as follows from Eq.~(\ref{eq:app}). Their numerical values are $1.9\times10^{-9}$ and $8.2\times10^{-8}$, respectively.

Finally, we express the Yukawa interactions in terms of the physical Higgs bosons and fermion mass eigenstates:
\begin{align}
\mathcal{L}_Y^{\rm int} 
&= -\bar{u}(V_{\rm CKM}\, \rho_dP_R - \rho_u^\dagger V_{\rm CKM}P_L )dH^+  -\bar{\nu}\rho_e P_R e H^+ +\text{h.c.} \notag\\
& -\sum^{3}_{a =1} \bar{f} \left[R_{1a}\frac{M_f^{\rm diag}}{v} + \frac{R_{2a}}{\sqrt{2}}(\rho_f P_R + \rho_f^\dagger P_L) - 2iI_f\frac{R_{3a}}{\sqrt{2}}(\rho_fP_R - \rho_f^\dagger P_L)   \right]H_af, 
\label{Yukawa3}
\end{align}
where $H_1=h$, $P_L$  $(P_R)$ is the left- (right-)handed chirality projection operators, and $R$ is the orthogonal matrix that diagonalizes the neutral Higgs squared-mass matrix, as defined in Eq.~(\ref{eq:rot}).
As discussed above, $\rho_d$ and $\rho_e$ are flavor diagonal due to the partial Yukawa alignment imposed in Eq.~(\ref{eq:rho_dee}), while $\rho_u$ generally contains flavor-violating elements.
For convenience, we use the shorthand notation $V_{ij}\equiv(V_{\rm CKM})_{ij}$ and $\rho_{ij}\equiv(\rho_u)_{ij}$, where the indices denote the corresponding quark flavors. For example, $V_{cb}=(V_{\rm CKM})_{23}$ and $\rho_{ct}=(\rho_u)_{23}$.

In summary, the partial Yukawa alignment allows us to reproduce the observed KM phase while avoiding tree-level FCNCs in the down-type quark and charged-lepton sectors. At the same time, sizable flavor-violating Yukawa interactions involving the top and charm quarks can arise in the up-type sector. These interactions play a crucial role in the collider phenomenology of the additional Higgs bosons, as discussed in Sec.~\ref{sec:phenomenology}.

\section{Constraints \label{sec:constraints}}

The parameter space of the SCPV 2HDM is subject to various theoretical and experimental constraints. In this section, we summarize the constraints imposed in our numerical analysis.

\subsection{Theoretical Constraints}

\begin{itemize}
\item \textbf{Vacuum stability:} The Higgs potential must be bounded from below (BFB) to ensure vacuum stability. In the general 2HDM, the BFB conditions impose nontrivial constraints on the quartic couplings. In our analysis, we impose the following conditions~\cite{Deshpande:1977rw,Nie:1998yn,Kanemura:1999xf,Branco:2011iw}:
\begin{eqnarray}
&&\lambda_1 \ge 0, \quad
\lambda_2 \ge 0, \quad
\sqrt{\lambda_1\lambda_2}+\lambda_3 \ge 0, \quad
\sqrt{\lambda_1\lambda_2}+\lambda_3+\lambda_4-|\lambda_5| \ge 0,  \\
&&\frac{1}{2}(\lambda_1+\lambda_2)+\lambda_3+\lambda_4+\lambda_5
-2|\lambda_6+\lambda_7| \ge 0 .
\end{eqnarray}

\item \textbf{Perturbative unitarity:} We impose perturbative unitarity by considering all possible two-body scattering processes involving scalar bosons and longitudinally polarized electroweak gauge bosons in the high-energy limit. The corresponding $s$-wave scattering matrix can be decomposed into block-diagonal matrices~\cite{Ginzburg:2005dt,Kanemura:2015ska}. We require all its eigenvalues $a_i$ to satisfy
\begin{equation}
|a_i| < \frac{1}{2},
\end{equation}
which ensures that the tree-level scattering amplitudes remain within the perturbative unitarity bound.

\item \textbf{Perturbativity:} We further require the quartic couplings to remain within the perturbative regime by imposing $|\lambda_i|\leq4\pi$ for $i=1,\ldots,7$.

\end{itemize}

\subsection{Oblique parameters}

Electroweak precision observables provide important indirect probes of physics beyond the SM. New physics contributions to the electroweak gauge boson self-energies can be parameterized in terms of the oblique parameters $S$, $T$, and $U$~\cite{Peskin:1991sw,Peskin:1990zt}.
In extended Higgs sectors, including the general 2HDM, additional scalar bosons contribute to these parameters at the one-loop level. The relevant expressions are given in Refs.~\cite{Grimus:2007if,Grimus:2008nb,Branco:2011iw}.
In our analysis, we impose constraints on the $S$ and $T$ parameters, neglecting $U$, whose contribution is generally suppressed in the parameter region of interest. We adopt the global-fit results~\cite{Asadi:2022xiy}
\begin{align}
S &= S_{0} \pm \sigma_{S} = 0.03 \pm 0.08, \\
T &= T_{0} \pm \sigma_{T} = 0.07 \pm 0.06 ,
\end{align}
under the assumption of $U=0$ and with the correlation coefficient $\rho_{ST}=0.916$.
To quantify the agreement between the model predictions and electroweak precision data, we define
\begin{equation}
\chi^{2} =
\frac{(S-S_{0})^{2}}{\sigma_{S}^{2}(1-\rho_{ST}^{2})}
+
\frac{(T-T_{0})^{2}}{\sigma_{T}^{2}(1-\rho_{ST}^{2})}
-
\frac{2\rho_{ST}^{} (S-S_{0})(T-T_{0})}
{\sigma_{S}\sigma_{T}(1-\rho_{ST}^{2})}.
\end{equation}
We require
\begin{equation}
\Delta\chi^{2} = \chi^{2}-\chi^{2}_{\rm min} < 5.991 ,
\end{equation}
corresponding to the 95\% confidence region for two degrees of freedom.

The Yukawa interactions in the SCPV 2HDM give rise to additional contributions to flavor observables through charged and neutral Higgs boson exchange. In this work, we impose constraints from $B_d$–$\bar B_d$ mixing, $B_s$–$\bar B_s$ mixing, $D^0$–$\bar D^0$ mixing, and the radiative decay $b\to s\gamma$. These observables provide stringent indirect constraints on extended Higgs sectors and have been extensively studied in the context of general 2HDMs~\cite{Branco:2011iw,Buras:2010pz,Crivellin:2013wna}.

\subsection{Flavor physics constraints}
\begin{itemize}
 \item \textbf{$B_q$–$\bar B_q$ mixing:} Neutral $B_q$-meson mixing ($q=d,s$) is induced by box diagrams involving charged-current interactions. In the SM, the dominant contribution arises from box diagrams with internal top quarks and $W$ bosons~\cite{Inami:1980fz,Buras:2001ra}. In the general 2HDM, additional contributions arise from box diagrams involving one or two charged Higgs bosons~\cite{Geng:1988bq,Urban:1997gw,Crivellin:2013wna,Branco:2011iw,Altunkaynak:2015esa}.
The SM contribution is expressed in terms of the loop function
\begin{equation}
I_{WW} =
1+\frac{9}{1-y_W}-\frac{6}{(1-y_W)^2}
-\frac{6}{y_W}\left(\frac{y_W}{1-y_W}\right)^3\ln y_W ,
\qquad
y_W=\frac{m_t^2}{m_W^2}.
\end{equation}
The additional contributions from mixed $W$–$H^\pm$ and charged-Higgs box diagrams can be written as
\begin{align}
I_{HW}^{(q)} &=
\left(
\frac{\rho_{tt}^\ast}{y_t^{\rm SM}}
+\frac{V_{cb}\,\rho_{ct}^\ast}{V_{tb}\,y_t^{\rm SM}}
\right)
\left(
\frac{\rho_{tt}}{y_t^{\rm SM}}
+\frac{V_{cq}^\ast\,\rho_{ct}}{V_{tq}^\ast\,y_t^{\rm SM}}
\right)
y_H F_{HW}(x,y_H), \\
I_{HH}^{(q)} &=
\left(
\frac{\rho_{tt}^\ast}{y_t^{\rm SM}}
+\frac{V_{cb}\,\rho_{ct}^\ast}{V_{tb}\,y_t^{\rm SM}}
\right)^2
\left(
\frac{\rho_{tt}}{y_t^{\rm SM}}
+\frac{V_{cq}^\ast\,\rho_{ct}}{V_{tq}^\ast\,y_t^{\rm SM}}
\right)^2
y_H F_{HH}(y_H),
\end{align}
where
\begin{equation}
y_H=\frac{m_t^2}{m_{H^\pm}^2},
\qquad
x=\frac{m_{H^\pm}^2}{m_W^2}\quad \textrm{and} \quad y_t^{\rm SM} = \frac{\sqrt2 m_t}{v},
\end{equation}
and the loop functions are given by
\begin{align}
F_{HW}(x,y_H) &=
\frac{(2x-8)\ln y_H}{(1-x)(1-y_H)^2}
+
\frac{6x\ln y_W}{(1-x)(1-y_W)^2}
-
\frac{8-2y_W}{(1-y_W)(1-y_H)}, \\
F_{HH}(y_H) &=
\frac{1+y_H}{(1-y_H)^2}
+
\frac{2y_H\ln y_H}{(1-y_H)^3}.
\end{align}
To compare the predicted mixing amplitudes with experimental constraints, we adopt the UTfit parameterization~\cite{UTfit}:
\begin{equation}
M_{12}^{(q)} = M_{12}^{\rm SM}\,
C_{B_q} e^{2 i \phi_{B_q}},
\end{equation}
The new physics contributions are then expressed as
\begin{equation}
C_{B_q} e^{2 i \phi_{B_q}}
=
1+\frac{I_{HW}^{(q)}+I_{HH}^{(q)}}{I_{WW}} .
\end{equation}
We require the predicted values of $C_{B_d}$, $C_{B_s}$, and the corresponding CP-violating phases to be consistent with the allowed regions obtained from global flavor fits~\cite{Bona:2007vi,Charles:2004jd,UTfit,CKMfitter}. We use the following constraints:
\begin{align}
C_{B_d} &= 1.09 \pm 0.18 , &
\phi_{B_d} &= (-1.8 \pm 3.8)^\circ , \\
C_{B_s} &= 1.10 \pm 0.12 , &
\phi_{B_s} &= (-0.1 \pm 1.0)^\circ ,
\end{align}
These constraints are imposed in our numerical parameter scan.

\item \textbf{$D^0-\bar D^0$ mixing: } In the general 2HDM, $D^0$–$\bar D^0$ mixing receives tree-level contributions from the exchange of flavor-changing neutral Higgs bosons. These contributions depend on $\rho_{uc}$ and $\rho_{cu}$ and are suppressed by the squared masses of the exchanged neutral scalars.
Additional contributions arise from box diagrams involving Higgs bosons and quarks, which can constrain combinations of other flavor-violating up-type Yukawa couplings, including $\rho_{ut}$ and $\rho_{ct}$. Thus, $D^0$–$\bar D^0$ mixing provides an important probe of flavor violation in the up-type quark sector.
Given the sizable long-distance hadronic uncertainties in the SM prediction~\cite{Golowich:2007ka,Isidori:2010kg}, we adopt a conservative constraint by requiring the new physics contribution to the mixing amplitude not to exceed the experimentally measured value.

\item \textbf{Radiative decay $b\to s\gamma$: }The inclusive radiative decay $B\to X_s\gamma$ provides one of the most stringent constraints on charged Higgs bosons in extended Higgs sectors~\cite{Misiak:2015xwa,Misiak:2020vlo,Crivellin:2019dun}.

The effective Hamiltonian describing the $b\to s\gamma$ transition is given by~\cite{Buras:2001ra}
\begin{equation}
\mathcal{H}_{\rm eff}
=
-\frac{4 G_F}{\sqrt{2}} V_{tb}V_{ts}^*
\sum_i C_i(\mu) O_i(\mu),
\end{equation}
where the relevant electromagnetic and chromomagnetic dipole operators are defined as
\begin{align}
O_7 &=
\frac{e}{16\pi^2} m_b
(\bar s_L \sigma^{\mu\nu} b_R)F_{\mu\nu}, \\
O_8 &=
\frac{g_s}{16\pi^2} m_b
(\bar s_L \sigma^{\mu\nu} T^a b_R)G_{\mu\nu}^a .
\end{align}

In the SCPV 2HDM, the Wilson coefficients receive additional contributions from charged Higgs loops involving up-type quarks. We express the coefficients as
\begin{equation}
C_{7,8}^{\rm eff}
=
C_{7,8}^{\rm SM}
+
\Delta C_{7,8},
\end{equation}
where $\Delta C_{7,8}$ denote the new physics contributions, which depend on the charged Higgs mass and the Yukawa coupling matrices. Both top- and charm-quark loop contributions are included in our numerical analysis.
%
To evaluate the branching ratio, we use the numerical approximation based on the NNLO SM calculation and retain only terms linear in the new physics contributions~\cite{Misiak:2015xwa,Misiak:2020vlo,Crivellin:2019dun}:
\begin{equation}
\frac{\mathrm{BR}(B\to X_s\gamma)_{\rm 2HDM}}
{\mathrm{BR}(B\to X_s\gamma)_{\rm SM}}
=
1+\frac{\Delta \mathrm{BR}_{\rm 2HDM}}{\mathrm{BR}(B\to X_s\gamma)_{\rm SM}}.
\end{equation}
We compare the resulting prediction with the experimental world average and the NNLO SM prediction~\cite{Crivellin:2019dun}:
\begin{equation}
\Delta \mathrm{BR}_{\rm 2HDM}\times 10^{4}=
-8.22\,\Re(\Delta C_7) -1.99\,\Re(\Delta C_8) .
\end{equation}

In our numerical analysis, we require the predicted branching ratio to be consistent with the experimental measurement within the $2\sigma$ uncertainty range.
%

\end{itemize}

\subsection{Electric dipole moments}

EDMs provide stringent constraints on new sources of CPV. In particular, the electron EDM (eEDM) is subject to the upper bound $|d_e|<4\times10^{-30}\,e\,\mathrm{cm}$~\cite{Roussy:2022cmp}.
In our scenario, the relative phase $\xi$ of the Higgs VEVs is the fundamental source of CPV. The dominant contributions to the eEDM arise from two-loop Barr–Zee (BZ) diagrams~\cite{Barr:1990vd}. In the Higgs alignment limit, the BZ contributions can be decomposed as
\begin{align}
d_e^{\rm BZ} = \sum_{V=\gamma,Z,W}[d_e^V(S) + d_e^V(t)], 
\end{align}
where $d_e^V(S)$ and $d_e^V(t)$ denote the contributions from scalar and top-quark loops, respectively. Here, $V=\gamma,Z,W$ labels the gauge boson connecting the loop subdiagram to the external electron line.
The individual contributions are given by~\cite{Kanemura:2020ibp}
\begin{align}
d_e^{V^0}(S)&= 
-\left(\frac{1}{16\pi^2}\right)^2\frac{4em_e}{v}
\sum_{\phi  = H_2,H_3}
\lambda_{\phi H^+H^-} g_{H^+H^-V^0}^{}\,g_{V^0ee}^{}\rho_e^{\phi }\notag\\
&\quad\quad \int_0^1dx x F\left(\frac{m_{H^\pm}^2}{x(1-x)} , m_{V^0}^2,m_{\phi}^2\right), \\
d_e^W(S) &=  \left(\frac{1}{16\pi^2}\right)^2\frac{em_e}{v} 
\sum_{\phi  = H_2,H_3}
g\text{Im}[\zeta_e^*g_{H^+\phi W}\lambda_{\phi H^+ H^-}] \int_0^1dx x F\left(\frac{m_{H^\pm}^2}{1-x} + \frac{m_{\phi}^2}{x}, m_W^2,m_{H^\pm}^2\right), \\
d_e^{V^0}(t) &= \sum_{\phi  = H_2,H_3}\left(\frac{1}{16\pi^2}\right)^2\frac{16em_e}{v}g_{V^0ee}^{}g_{V^0tt}^{} \frac{m_t^{2}}{v}\notag\\
&\int_0^1dx \left(\frac{2x^2-2x+1}{1-x}\rho_e^{\phi }\kappa_t^{\phi } + \frac{\kappa_e^{\phi }\rho_t^{\phi }}{1-x}\right) F\left(\frac{m_t^2}{x(1-x)} , m_V^2,m_{\phi}^2\right), \\
d_e^W(t) &= 3\left(\frac{1}{16\pi^2}\right)^2\frac{em_e}{v} \frac{g^2}{v}\int_0^1 dx\, \text{Im}\Big\{  \zeta_e^*\left[x m_b^2 \zeta_d + (2-x)m_t^2 \zeta_u^* \right] \Big\}\notag\\
& \left(Q_d + \frac{1-x}{x}Q_u \right) F\left(\frac{m_{b}^2}{1-x} + \frac{m_{t}^2}{x}, m_W^2,m_{H^\pm}^2 \right), 
\end{align}
where the relevant couplings and loop functions are defined as
\begin{align}
g_{H^+H^-V^0}^{} &= 
\begin{cases}
e ~~(V^0=\gamma)\\
\frac{g}{2c_W^{}}(c_W^2-s_W^2)
~~(V^0 = Z)
\end{cases},\quad 
g_{V^0ff}^{} = 
\begin{cases}
Q_f\,e ~~(V^0 = \gamma)\\
\frac{g}{c_W^{}}\left(I_f - Q_f\,s_W^2\right)~~(V^0 = Z)
\end{cases}, \\
g_{H^+\phi  W}^{} &= 
\begin{cases}
\frac{g}{2}(c_\alpha + i s_\alpha) ~~(\phi =H_2)\\
\frac{g}{2}(-s_\alpha - i c_\alpha) ~~(\phi =H_3)
\end{cases}, \\
\begin{pmatrix}
\kappa_e^{H_2}\\
\kappa_e^{H_3}
\end{pmatrix}
&=\begin{pmatrix}
c_\alpha & -s_\alpha \\
-s_\alpha & -c_\alpha
\end{pmatrix}
\begin{pmatrix}
\Re (\zeta_e)\\
\Im (\zeta_e)
\end{pmatrix},\quad \quad \quad 
\begin{pmatrix}
\rho_e^{H_2}\\
\rho_e^{H_3}
\end{pmatrix}
=\begin{pmatrix}
s_\alpha & c_\alpha \\
c_\alpha & -s_\alpha
\end{pmatrix}
\begin{pmatrix}
\Re (\zeta_e)\\
\Im (\zeta_e)
\end{pmatrix}, \\
\begin{pmatrix}
\kappa_t^{H_2}\\
\kappa_t^{H_3}
\end{pmatrix}
&=\begin{pmatrix}
c_\alpha & s_\alpha \\
-s_\alpha & c_\alpha
\end{pmatrix}
\begin{pmatrix}
\Re (\zeta_t)\\
\Im (\zeta_t)
\end{pmatrix},\quad \quad \quad 
\begin{pmatrix}
\rho_t^{H_2}\\
\rho_t^{H_3}
\end{pmatrix}
=\begin{pmatrix}
-s_\alpha & c_\alpha \\
-c_\alpha & -s_\alpha
\end{pmatrix}
\begin{pmatrix}
\Re (\zeta_t)\\
\Im (\zeta_t)
\end{pmatrix}, 
\end{align}
and 
\begin{align}
F(m_1^2,m_2^2,m_3^2) & = -\int_0^{\infty}dy
\frac{y}{(y+m_1^2)(y+m_2^2)(y+m_3^2)}.
\end{align}
The $\zeta_t$ parameter is defined as $\zeta_t = v\rho_{tt}/(\sqrt{2}m_t) = \rho_{tt}/y_t^{\rm SM}$ and $\zeta_e$ is defined in Eq.~\ref{eq:zeta-d-e}.

\begin{figure}[h!]
    \centering
    \includegraphics[width=0.6\linewidth]{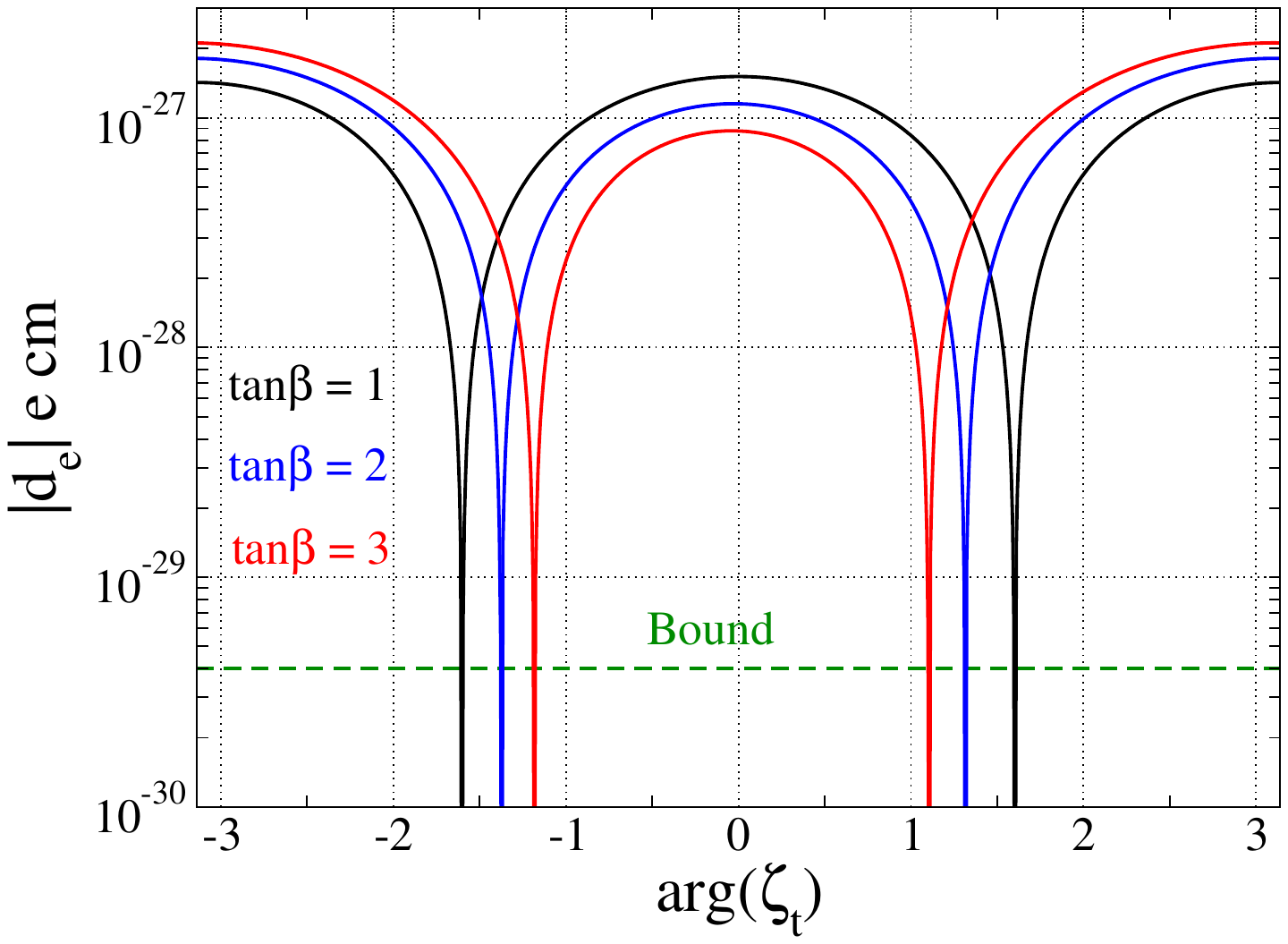}
    \caption{
    Magnitude of the electron EDM, $|d_e|$, as a function of $\arg(\zeta_t)$ for $\tan\beta=1$ (black), $2$ (blue), and $3$ (red). We take $m_{H^\pm}=m_{H_2}=m_{H_3}=300$ GeV, $|\zeta_t|=0.1$, $\tilde{\zeta}_e=1$, $\tilde{\zeta}_d=50$, $\tan\xi=20$, and $\alpha=0.1$.}
    \label{fig:de}
\end{figure}

Fig.~\ref{fig:de} shows the dependence of $|d_e|$ on $\arg(\zeta_t)$ for $\tan\beta=1$ (black), $2$ (blue), and $3$ (red). For generic values of the CP-violating phase, the predicted eEDM is typically of order $10^{-27}\,e\,\mathrm{cm}$, exceeding the current experimental upper bound.
However, destructive interference between the scalar-loop and top-quark-loop contributions can significantly suppress the total eEDM. Such cancellations can be realized by appropriately choosing the parameters of the unitary matrix $U_u$, which enters the up-type Yukawa coupling matrix $\rho_u$ in Eq.~(\ref{eq:rho_u}).
We also consider constraints from other EDM measurements. In particular, the neutron EDM is bounded by $|d_n|<1.8\times10^{-26}\,e\,\mathrm{cm}$~\cite{Abel:2020pzs}. In the parameter region considered here, the predicted neutron EDM is typically about one order of magnitude below this bound, making the eEDM constraint more restrictive.

\subsection{Loop-Induced Decays of the Discovered Higgs Boson} \label{sec:loop-induced}

\begin{figure}[t!]
    \centering
    \includegraphics[width=0.6\linewidth]{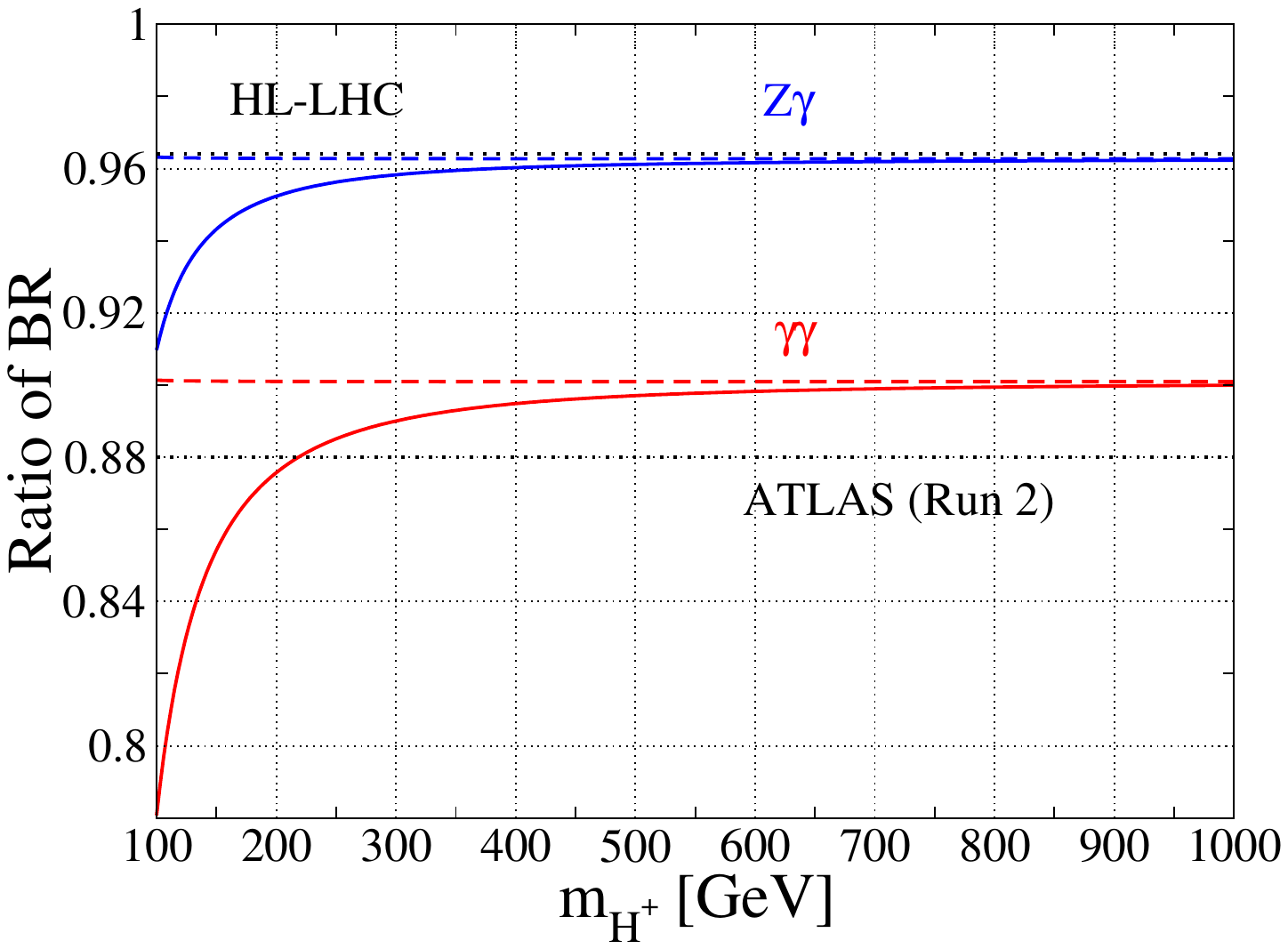}
    \caption{BRs of $h \to \gamma\gamma$ (red) and $h \to Z\gamma$ (blue) normalized by the corresponding SM predictions as a function of $m_{H^\pm}$ in the Higgs alignment limit. The dashed curves denote the asymptotic values as $m_{H^\pm} \to \infty$. The dotted lines indicate the 95\% CL 
    lower limits on the $h \to \gamma\gamma$ decay BR from the ATLAS experiment and the HL-LHC projection, where the latter assumes the central value matches the SM prediction.
    }
    \label{fig:hgaga}
\end{figure}

A characteristic feature of the SCPV 2HDM is the non-decoupling nature of the additional Higgs bosons. In particular, charged Higgs boson loops induce nonvanishing corrections to the decay rates of the SM-like Higgs boson into $\gamma\gamma$ and $Z\gamma$, even in the Higgs alignment limit.
The partial decay widths for $h\to\gamma\gamma$ and $h\to Z\gamma$ are given by~\cite{Gunion:1989we}
\begin{align}
\Gamma(h \to \gamma\gamma)&=  \frac{\sqrt{2}G_F\alpha_{\text{em}}^2m_h^3}{16\pi^3 }  \left|(t,W)_{\gamma\gamma}^{\rm SM} + \frac{v\lambda_{H^+H^-h}}{2m_h^2}\left[1 - \frac{1}{\tau_h} \arcsin^2(\sqrt{\tau_h})\right]
\right|^2,\\
\Gamma(h \to Z\gamma)&=  \frac{\sqrt{2}G_F\alpha_{\text{em}}^2m_h^3}{8\pi^3 } 
\left(1 - \frac{m_Z^2}{m_h^2}\right)^3
\Bigg|(t,W)_{Z\gamma}^{\rm SM} + \frac{c_W^2-s_W^2}{4s_Wc_W}\frac{v\lambda_{H^+H^-h}}{m_h^2-m_Z^2} \notag\\
&\Bigg\{1 -
\frac{4m_{H^\pm}^2}{m_h^2-m_Z^2}
\left[\arcsin^2(\sqrt{\tau_h})-\arcsin^2(\sqrt{\tau_Z})
\right]\notag\\
&+\frac{2m_Z^2}{m_h^2-m_Z^2}\left[\beta_Z\arctan(\beta_Z^{-1}) -\beta_h\arctan(\beta_h^{-1})\right]
\Bigg\}
\Bigg|^2,
\end{align}
where $\beta_h=\sqrt{\tau_h^{-1}-1}$ and $\beta_Z=\sqrt{\tau_Z^{-1}-1}$, with $\tau_h=m_h^2/(4m_{H^\pm}^2)$ and $\tau_Z=m_Z^2/(4m_{H^\pm}^2)$. These expressions are valid for $\tau_h,\tau_Z\leq1$.
The SM contributions from the $W$-boson and top-quark loops are denoted by $(t,W)_{\gamma\gamma}^{\rm SM}$ and $(t,W)_{Z\gamma}^{\rm SM}$, whose numerical values are approximately $1.6$ and $2.9$, respectively. The trilinear coupling $\lambda_{H^+H^-h}$ is given in Eq.~\ref{eq:hphmh}.
In the limit $m_{H^\pm}^2\gg m_h^2,m_Z^2$, the decay widths approach
\begin{align}
\Gamma(h \to \gamma\gamma)&=  \frac{\sqrt{2}G_F\alpha_{\text{em}}^2m_h^3}{16\pi^3 }  \left|(t,W)_{\gamma\gamma}^{\rm SM} - \frac{1}{12 } + {\cal O}(\tau_h)\right|^2,   \label{eq:hgg_asym}\\
\Gamma(h \to Z\gamma)&=  \frac{\sqrt{2}G_F\alpha_{\text{em}}^2m_h^3}{8\pi^3 } 
\left(1 - \frac{m_Z^2}{m_h^2}\right)^3
\Bigg|(t,W)_{Z\gamma}^{\rm SM} - \frac{c_W^2-s_W^2}{24s_Wc_W} + {\cal O}(\tau_h,\tau_Z)\Bigg|^2.\label{eq:hzg_asym}
\end{align}
These expressions explicitly demonstrate the non-decoupling behavior of the charged Higgs contribution. Although the charged Higgs boson becomes heavy, its trilinear coupling to $h$ grows proportionally to $m_{H^\pm}^2$. Consequently, the charged Higgs loop contribution approaches a nonzero constant rather than vanishing in the large-mass limit.
Furthermore, the charged Higgs contribution interferes destructively with the SM amplitudes, leading to a suppression of both decay rates relative to their SM predictions.

Fig.~\ref{fig:hgaga} shows the branching ratios of $h\to\gamma\gamma$ and $h\to Z\gamma$, normalized to their corresponding SM predictions, as functions of $m_{H^\pm}$. As expected from Eqs.~(\ref{eq:hgg_asym}) and (\ref{eq:hzg_asym}), the deviations remain finite even in the large charged Higgs mass limit.
The Higgs signal strength for the $\gamma\gamma$ channel $\mu_{\gamma\gamma}$ has been constrained by the ATLAS~\cite{ATLAS:2022jtk,ATLAS:2022vkf,ATLAS:2022tnm} and CMS~\cite{CMS:2022dwd} Collaborations. The quoted 95\% confidence intervals are
\begin{align}
0.88 \leq \mu_{\gamma\gamma} &\leq 1.18,\quad  \textrm{[ATLAS]} \\
0.94 \leq \mu_{\gamma\gamma} &\leq 1.26,\quad  \textrm{[CMS]}. 
\end{align}
In the Higgs alignment limit, $\mu_{\gamma\gamma}$ is determined by the ratio of the branching ratios of $h\to \gamma\gamma$. 
The ATLAS constraint is indicated in Fig.~\ref{fig:hgaga} and corresponds to a lower bound on the charged Higgs mass of approximately $200$ GeV under the assumptions adopted in this analysis. The quoted CMS interval, on the other hand, is in tension with the predicted suppression of the diphoton decay rate.
At the HL-LHC, the precision on $\kappa_\gamma$ is projected to reach approximately $1.9\%$ at the $1\sigma$ level~\cite{deBlas:2019rxi}. Assuming that the measured central value agrees with the SM prediction, the corresponding projected 95\% confidence limit is also shown in Fig.~\ref{fig:hgaga}. The HL-LHC will therefore provide a powerful test of the characteristic non-decoupling effects predicted in this scenario.

\subsection{LHC direct searches}

Direct searches for additional Higgs bosons at the LHC provide important constraints on the Yukawa interactions of the SCPV 2HDM. In particular, the production and decay rates of the additional Higgs bosons are sensitive to the up-type Yukawa couplings $\rho_{tu}$, $\rho_{tc}$, and $\rho_{tt}$.

\begin{itemize}
\item {\bf Constraints on $\rho_{tu}$ and $\rho_{tc}$: }
The flavor-violating Yukawa couplings $\rho_{tu}$ and $\rho_{tc}$ induce processes of the form
\begin{equation}
pp \to t\,\phi \to t (t \bar q),
\qquad q = u, c,
\end{equation}
where $\phi=H_2,H_3$ and $q=u,c$. These processes can produce same-sign top-quark pairs, leading to characteristic same-sign dilepton signatures.
Such signatures have been searched for by the ATLAS and CMS Collaborations~\cite{CMS:2023xpx,ATLAS:2023tlp}, providing direct probes of flavor-changing Yukawa interactions in extended Higgs sectors~\cite{Altunkaynak:2015esa,Crivellin:2013wna}.
The absence of a significant excess over the SM background places constraints on the flavor-violating couplings. For additional Higgs boson masses around $300$–$400$ GeV, couplings of order $0.2$–$0.3$ can be probed, while couplings of order unity can be constrained for scalar masses approaching the TeV scale.
The precise bounds depend on the Higgs mass spectrum, decay branching ratios, and possible interference effects between the neutral Higgs bosons. These effects are particularly relevant in the SCPV 2HDM, where the additional Higgs bosons can have correlated masses and Yukawa interactions.

\item {\bf Constraints on $\rho_{tt}$: } The coupling $\rho_{tt}$ can be probed through charged Higgs boson production in association with top and bottom quarks:
\begin{equation}
pp \to tbH^\pm \to tb\,tb,
\end{equation}
Searches by the ATLAS Collaboration~\cite{ATLAS:2021upq,ATLAS:2023tlp,Hou:2025tjp} constrain the product $\sigma(pp\to tbH^\pm)\times\mathrm{BR}(H^\pm\to tb)$ over a wide range of charged Higgs masses.
These results can be translated into constraints on $\rho_{tt}$, depending on the charged Higgs mass and the relevant decay branching ratios. In particular, couplings of order unity are strongly constrained for charged Higgs boson masses below the TeV scale.
The constraints on the Yukawa couplings are generally correlated. For instance, the limits on $\rho_{tu}$ and $\rho_{tc}$ depend on $\rho_{tt}$ through the production and decay rates of the additional Higgs bosons. Moreover, interference between $H_2$ and $H_3$ can modify the predicted collider signals.
In our numerical analysis, we impose the relevant LHC search constraints by restricting the Yukawa couplings and Higgs boson masses to the experimentally allowed parameter regions.

\begin{figure}[t!]
    \centering
    \includegraphics[width=0.6\linewidth]{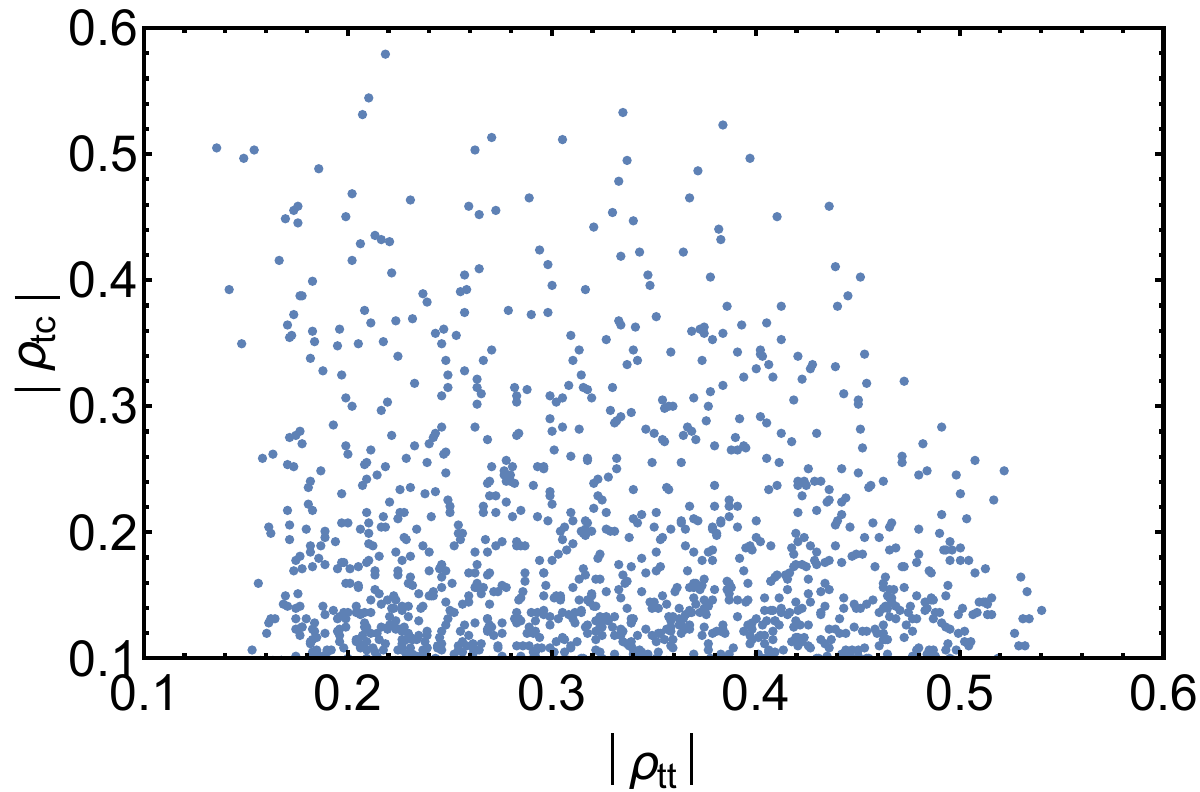}
    \caption{Allowed values of $|\rho_{tt}|$ and $|\rho_{tc}|$ after imposing the theoretical and experimental constraints discussed in this section.}
    \label{fig:rtt-rtc}
\end{figure}
\end{itemize}

Combining the theoretical and experimental constraints discussed in this section, we obtain the allowed parameter space shown in Fig.~\ref{fig:rtt-rtc}, where the surviving values of $\rho_{tt}$ and $\rho_{tc}$ are displayed.

\section{Collider Phenomenology}
\label{sec:phenomenology}

We investigate the collider phenomenology of the SCPV 2HDM at the LHC. Throughout this analysis, we work in the Higgs alignment limit, where the production rates of the SM-like Higgs boson coincide with their SM predictions. Its decay rates also agree with the SM predictions at leading order, while the loop-induced $h\to\gamma\gamma$ and $h\to Z\gamma$ decays receive non-decoupling contributions from the charged Higgs boson, as discussed in Sec.~\ref{sec:loop-induced}.
We therefore focus on the production and decay of the additional Higgs bosons, $H_2$, $H_3$, and $H^\pm$, and explore their characteristic collider signatures.

\subsection{Production of the Additional Higgs Bosons}
\label{subsec:production}

The collider phenomenology of the additional Higgs bosons is strongly influenced by the flavor structure of the Yukawa coupling matrix $\rho_u$. As discussed in the previous sections, a sizable off-diagonal coupling $\rho_{tc}$ induces flavor-changing interactions between the top and charm quarks, giving rise to distinctive signatures involving multiple top quarks.

\begin{table}[t]
\begin{center}
\renewcommand{\arraystretch}{1.25}
\begin{tabular}{|c|c||c|c|}
\hline
\textbf{Parameter}  & \textbf{Range}
& \textbf{Parameter}    & \textbf{Range}  \\ \hline\hline
$m_{H^{\pm}},\,m_{H_{2}},\,m_{H_{3}}$ (GeV) & $200$--$800$
& $\alpha$         & $-\pi/2$--$\pi/2$ \\ \hline
$\tan\beta$         & $1$--$8$
& $\theta_{12},\,\theta_{23},\,\theta_{13}$ & $0$--$\pi$ \\ \hline
$\tan\xi$           & $0.5$--$50$
& $\delta_{12},\,\delta_{23},\,\delta_{13}$ & $0$--$2\pi$ \\ \hline
$\tilde{\zeta}_d$   & $0$--$50$
& $\tilde{\zeta}_e$ & $1$ \\ \hline
\end{tabular}
\caption{%
Input parameters and their scan ranges used in the phenomenological analysis. The additional Higgs boson masses $m_{H_2}$, $m_{H_3}$, and $m_{H^\pm}$ are scanned  keeping $m_{H_3}\geq m_{H_2}$, subject to the theoretical and experimental constraints. The angles $\theta_{ij}$ and phases $\delta_{ij}$ parameterize the unitary matrix $U_u$ associated with the right-handed up-type quarks, while $\alpha$ denotes the mixing angle between the additional neutral Higgs bosons. The charged-lepton alignment parameter $\tilde{\zeta}_e$ is fixed to unity throughout the analysis.
}
\label{tab:scan}
\end{center}
\end{table}

To identify phenomenologically viable parameter points, we perform a random scan over the parameter ranges listed in Table~\ref{tab:scan}. We retain only those points satisfying the theoretical and experimental constraints discussed in Sec.~\ref{sec:constraints}, including vacuum stability, perturbative unitarity, perturbativity,  electroweak precision observables, Higgs signal-strength measurements, flavor observables, EDM bounds, and direct searches for additional Higgs bosons.

\begin{itemize}
\item {\bf Neutral Higgs Production: } The production of the additional neutral Higgs bosons $H_2$ and $H_3$ is mainly controlled by the Yukawa couplings $\rho_{tt}$ and $\rho_{tc}$.
The diagonal coupling $\rho_{tt}$ governs gluon-fusion (ggF) production,
\begin{equation}
  gg \;\to\; H_{2,3}\,,
  \label{eq:ggF}
\end{equation}
which proceeds predominantly through a top-quark loop, similarly to the production of the SM Higgs boson.
The flavor-violating coupling $\rho_{tc}$, on the other hand, induces associated production with a top quark,
\begin{equation}
  c\,g \;\to\; t\,H_{2,3}\,.
  \label{eq:fcnc_prod}
\end{equation}
From Eq.~(\ref{Yukawa3}), the relevant interaction is given by
\begin{equation}
  \mathcal{L} \;\supset\;
  -\frac{\rho_{tc}}{\sqrt{2}}\sum_{i=2,3}
  \bar{t}\,P_R\,c
   \big(R_{2i} - i\,R_{3i}\big)H_i
  \;+\; \text{h.c.}
  \label{eq:lag_rhotc}
\end{equation}
The subsequent flavor-violating decays $H_{2,3}\to t\bar c+\bar t c$ lead to final states containing two top quarks. In particular, the same-sign top-quark final states $tt\bar c$ and $\bar t\bar t c$ provide distinctive signatures with relatively small SM backgrounds. Their leptonic decays give rise to same-sign dilepton events, which offer a promising probe of the flavor-violating Yukawa interactions.

\item {\bf Charged Higgs Production: } The production of charged Higgs bosons is controlled by their Yukawa interactions with quarks, particularly the $H^\pm tb$ and $H^\pm cb$ couplings. The relevant associated production processes include
\begin{align}
  g\,g  &\;\to\; H^+\,\bar{t}\,b \;+\; \text{h.c.}\,,
  \label{eq:Hpm_tb} \\
  g\,g  &\;\to\; H^+\,\bar{c}\,b \;+\; \text{h.c.}\,,
  \label{eq:Hpm_cb}
\end{align}
The latter process is sensitive to the flavor-violating coupling $\rho_{tc}$ and therefore provides a complementary probe of the Yukawa structure of the model.

\item {\bf Pair Production via Weak Gauge Bosons: }In addition to the Yukawa-induced production processes, the additional Higgs bosons can be pair-produced through Drell–Yan processes mediated by off-shell electroweak gauge bosons:
\begin{align}
  p\,p \;\to\; W^{\pm *}\,(Z^{*}) &\;\to\; H_2\,H^\pm\,(H_2\,H_3).
  \label{eq:DY_WH}
\end{align}
Unlike the Yukawa-induced processes, these production channels are determined by the electroweak gauge interactions and the scalar mixing structure. In the Higgs alignment limit, their couplings are fixed by the gauge symmetry.
For the parameter region considered here, their production cross sections are generally smaller than those of the dominant Yukawa-induced processes. Nevertheless, they can lead to characteristic multi-particle final states with relatively small SM backgrounds and thus provide complementary probes of the extended Higgs sector~\cite{Mondal:2021bxa}.

\end{itemize}

\subsection{Decay of the Additional Higgs Bosons}
\label{subsec:decays}

\begin{figure}[t]
\begin{center}
\includegraphics[width=7.4cm]{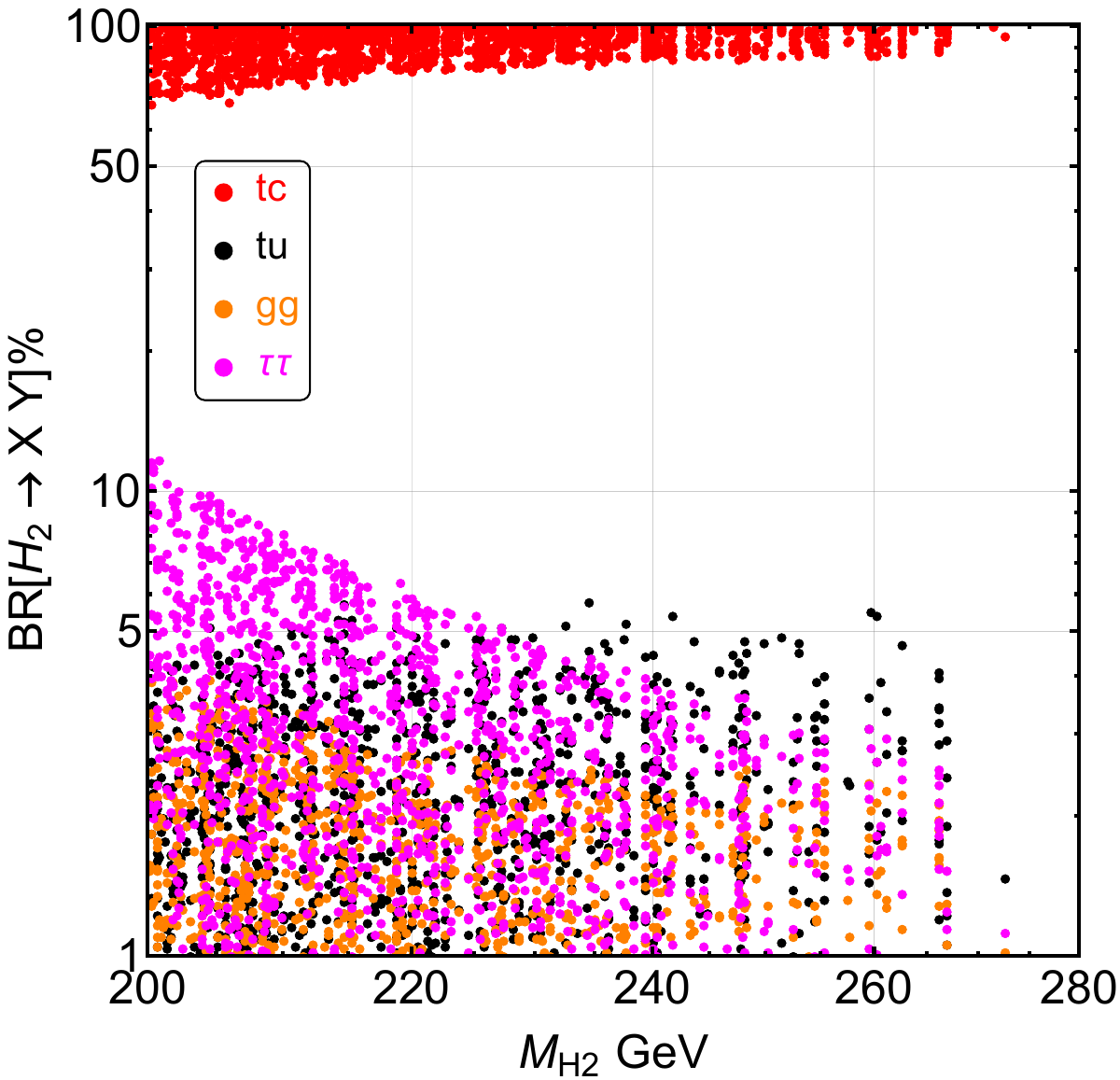}\hfill
\includegraphics[width=7.2cm]{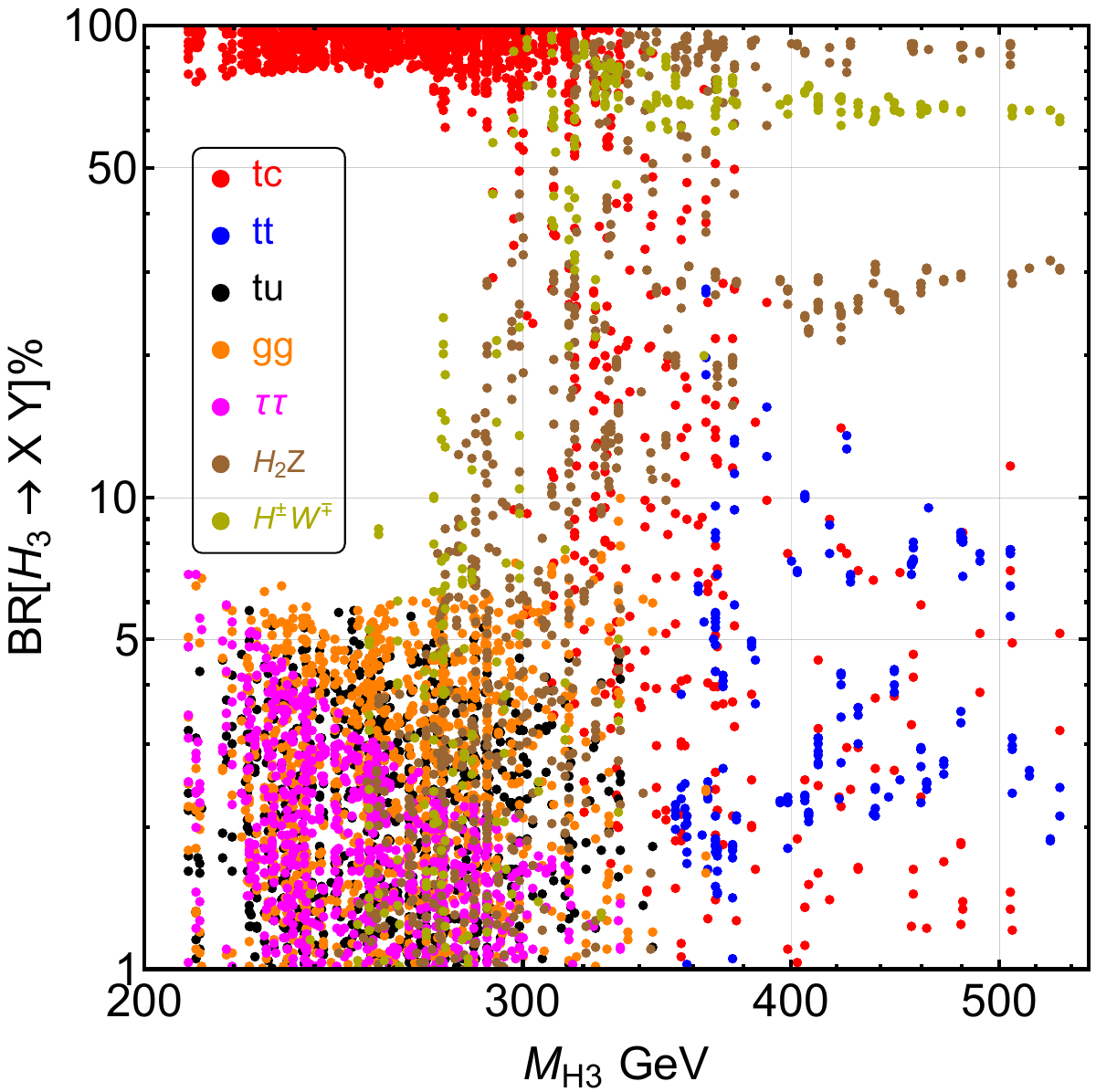}\\[0.5em]
\includegraphics[width=7.4cm]{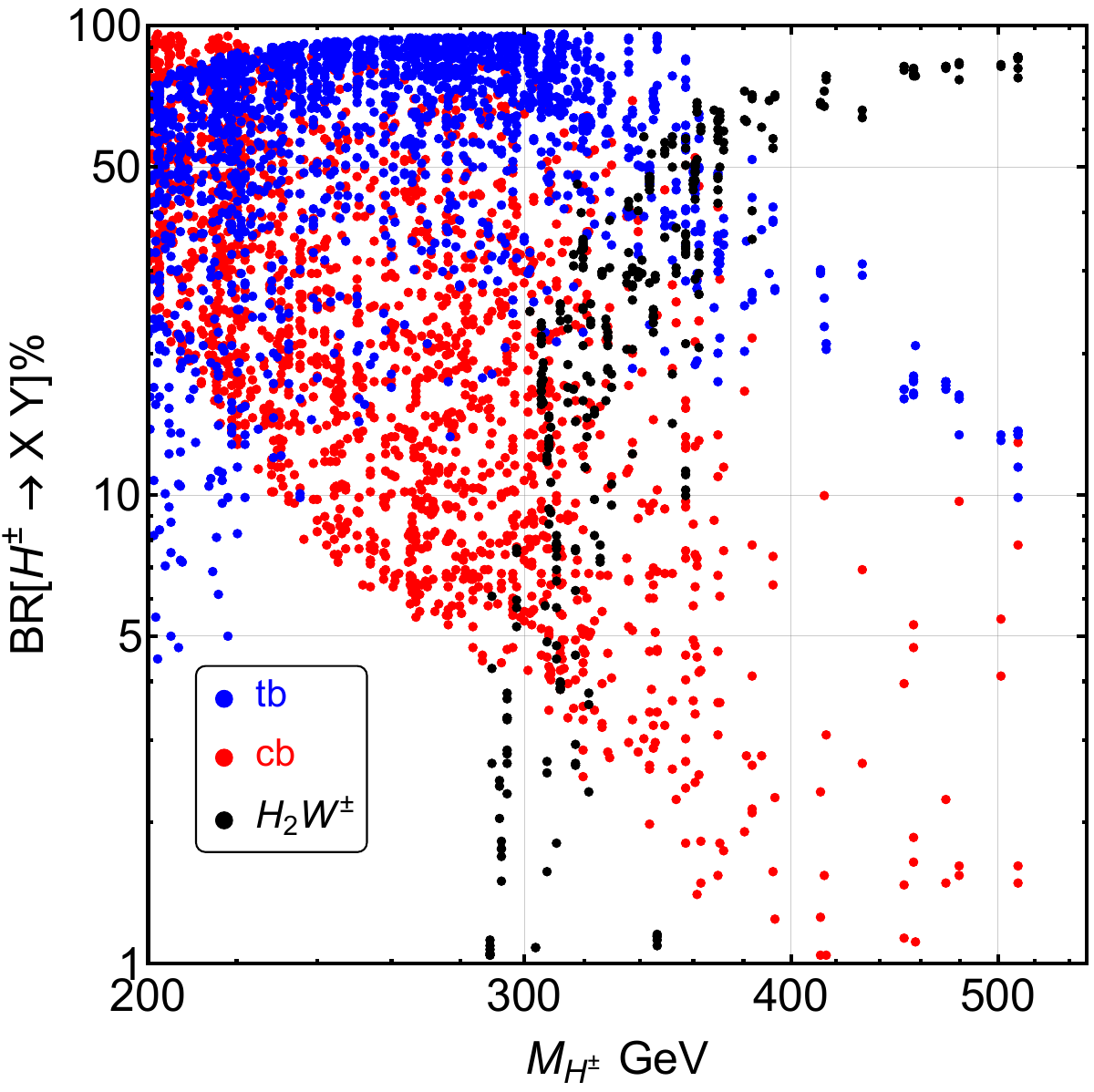}
\caption{%
  BRs of the BSM Higgs bosons as a function of their masses
  for all scan points satisfying the theoretical and experimental constraints.
  \textit{Top left:} $H_2$ decay modes.
  \textit{Top right:} $H_3$ decay modes.
  \textit{Bottom:} $H^\pm$ decay modes.
  The $y$-axis is shown on a logarithmic scale.
}
\label{fig:br}
\end{center}
\end{figure}

Fig.~\ref{fig:br} shows the branching ratios of the additional Higgs bosons as functions of their masses for the parameter points satisfying all the theoretical and experimental constraints discussed in Sec.~\ref{sec:constraints}.
The upper-left panel shows the branching ratios of $H_2$. The flavor-violating decay $H_2\to t\bar c+\bar t c$ is dominant, with a branching ratio ranging from approximately 80\% to nearly 100\%. The subdominant decay channels include $\tau^+\tau^-$, $gg$, and $t\bar u+\bar t u$, whose branching ratios are typically at the percent level or below. For the viable parameter points considered here, bosonic cascade decays such as $H_2\to H^\pm W^{\mp(*)}$ are suppressed, leaving the top-charm channel as the dominant decay mode.
The upper-right panel shows the branching ratios of the heavier neutral Higgs boson $H_3$. For relatively small $m_{H_3}$, the flavor-violating decay $H_3\to t\bar c+\bar t c$ dominates, with a branching ratio typically exceeding 80\%. The $\tau^+\tau^-$, $gg$, and top-up channels remain subdominant.
The decay pattern changes significantly once the bosonic channels $H_3\to H_2Z$ and $H_3\to H^\pm W^\mp$ become kinematically accessible. Depending on the mass spectrum, these cascade decays can dominate over the fermionic modes. Although $H_3\to t\bar t$ becomes allowed above the top-quark pair threshold, its branching ratio remains relatively small in the viable parameter region considered here.
The lower panel shows the branching ratios of the charged Higgs boson. The principal fermionic decay channels are $H^\pm\to tb$ and $H^\pm\to cb$. At relatively low charged Higgs masses, these two modes compete, with their relative importance determined by the Yukawa couplings. As the charged Higgs mass increases, the bosonic decay $H^\pm\to H_2W^\pm$ becomes kinematically accessible and can become the dominant decay mode.
The leptonic decay mode $H^\pm\to\tau^\pm\nu$ is subdominant, with its branching ratio remaining below $1\%$ throughout the parameter space considered in this work.

\begin{figure}[t]
\begin{center}
\includegraphics[width=7.4cm]{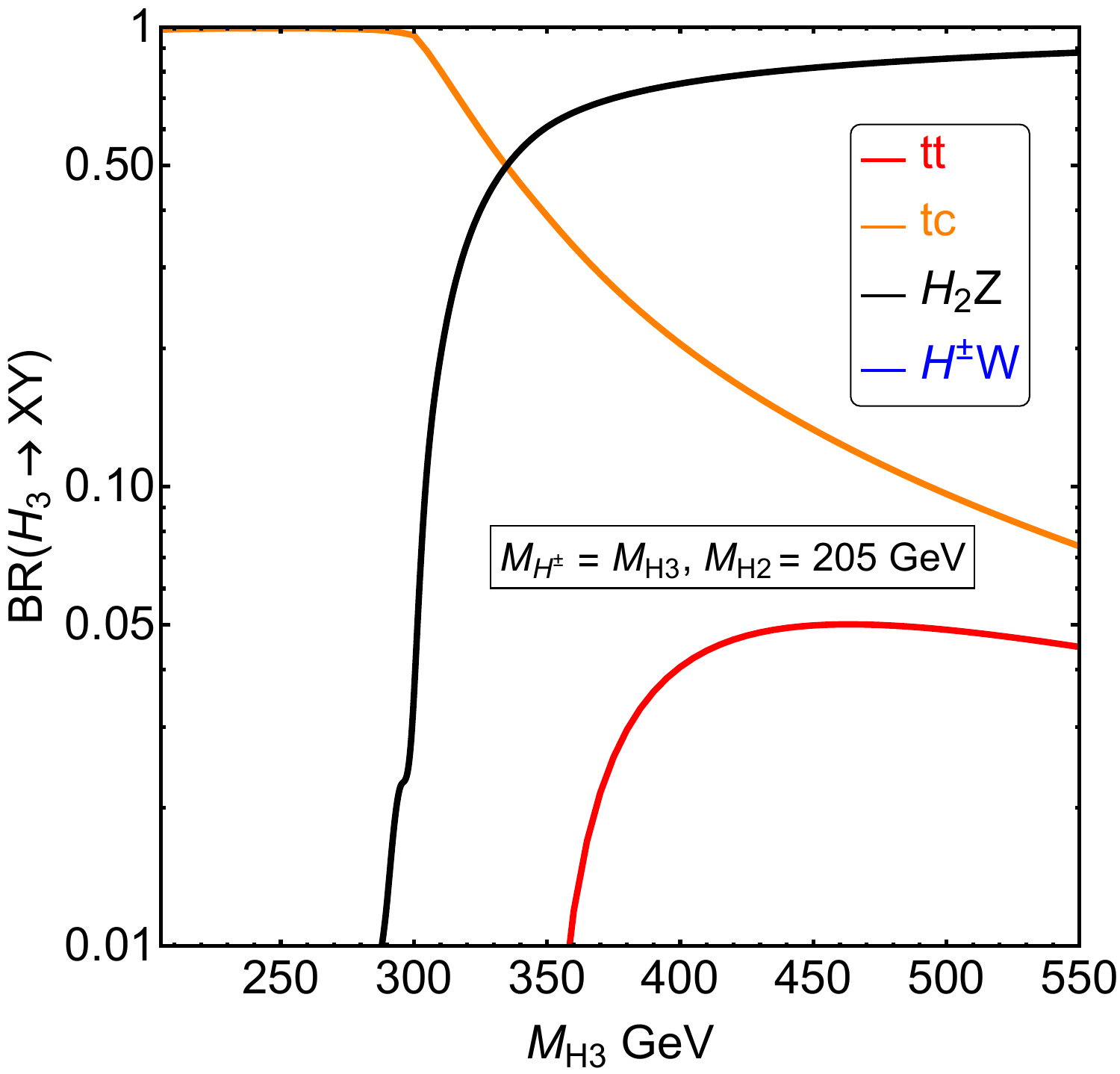}
\includegraphics[width=7.4cm]{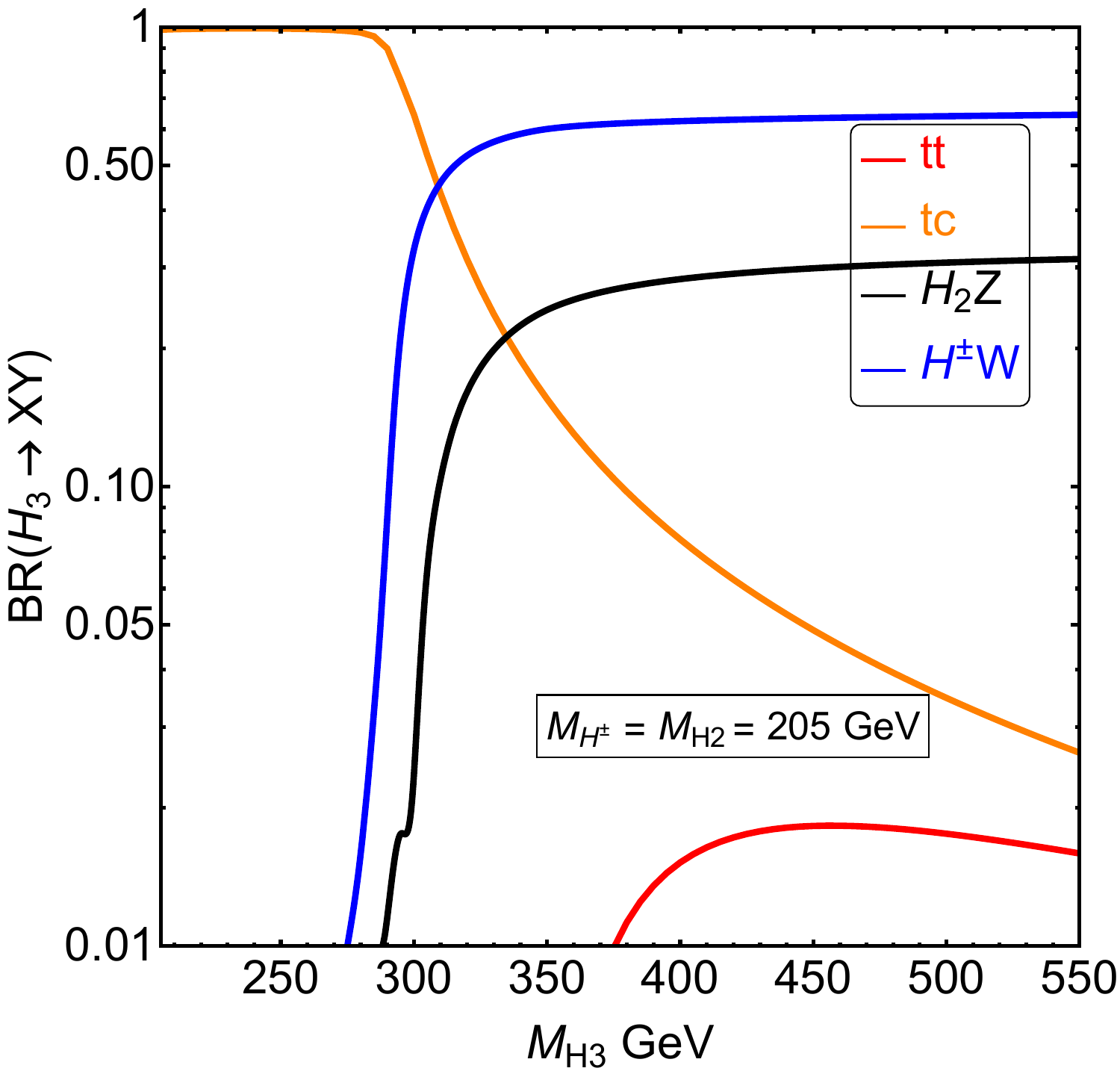}
\caption{%
Branching ratios of $H_3$ as functions of $m_{H_3}$ for two representative mass hierarchies. Left: $m_{H^\pm}=m_{H_3}$ and $m_{H_2}=205$ GeV, where the $H_3\to H^\pm W^\mp$ channel is kinematically forbidden. 
Right: $m_{H^\pm}=m_{H_2}=205$ GeV, where both $H_3\to H_2Z$ and $H_3\to H^\pm W^\mp$ become accessible above their respective thresholds. The remaining parameters are fixed to the values given in Table~\ref{tab:bp-decay}.
}
\label{fig:br-h3-mass}
\end{center}
\end{figure}

\begin{table}[t]
\centering
\renewcommand{\arraystretch}{1.15}
\setlength{\tabcolsep}{8pt}
\begin{tabular}{|c|c||c|c||c|c|}
\hline
\textbf{Parameter} & \textbf{Value}
& \textbf{Parameter} & \textbf{Value}
& \textbf{Parameter} & \textbf{Value} \\ \hline\hline
$\tan\beta$   & $2.88036$
& $\tan\xi$   & $40.95$
& $\alpha$    & $0.1$ \\ \hline
$\theta_{12}$ & $1.55147$
& $\theta_{23}$ & $1.43298$
& $\theta_{13}$ & $1.47311$ \\ \hline
$\delta_{12}$ & $6.20093$
& $\delta_{23}$ & $3.26784$
& $\delta_{13}$ & $1.60294$ \\ \hline
$\tilde{\zeta}_{d}$ & $50$
& $\tilde{\zeta}_{e}$ & $1$
&  &  \\ \hline
\end{tabular}
\caption{Values of various parameters fixed to explore the effect of mass hierarchy in the decay BR as shown in Fig.~\ref{fig:br-h3-mass}}
\label{tab:bp-decay}
\end{table}

To illustrate the dependence of the $H_3$ decay pattern on the additional Higgs mass spectrum, Fig.~\ref{fig:br-h3-mass} shows the branching ratios of $H_3$ as functions of $m_{H_3}$ for two representative mass hierarchies. The remaining model parameters are fixed to the values listed in Table~\ref{tab:bp-decay}.
In the left panel, we assume $m_{H^\pm}=m_{H_3}$ and fix $m_{H_2}=205$ GeV. In this configuration, the decay $H_3\to H^\pm W^\mp$ is kinematically forbidden throughout the mass range considered. Consequently, $H_3\to H_2Z$ is the only available two-body bosonic cascade decay and becomes dominant above its kinematic threshold, around $m_{H_3}\simeq 300$ GeV. Although the $t\bar t$ channel opens for $m_{H_3}>2m_t$, its branching ratio remains below approximately 5
In the right panel, we instead take $m_{H^\pm}=m_{H_2}=205$ GeV and vary $m_{H_3}$. In this case, both $H_3\to H_2Z$ and $H_3\to H^\pm W^\mp$ become kinematically accessible as $m_{H_3}$ increases. The charged-Higgs channel becomes dominant, reaching a branching ratio of approximately 60\%, while the $H_2Z$ channel contributes at the 25\%–30\% level.
These two benchmark configurations demonstrate that the relative importance of the bosonic decay channels is highly sensitive to the mass hierarchy among the additional Higgs bosons. In particular, the branching ratios of $H_3\to H_2Z$ and $H_3\to H^\pm W^\mp$ provide complementary information on the neutral and charged Higgs mass spectrum.


\subsection{Collider Signatures of the Neutral Higgs Bosons}

Combining the production mechanisms and decay patterns discussed above, the SCPV 2HDM predicts several characteristic collider signatures involving multiple top quarks and electroweak gauge bosons. These signatures provide complementary probes of the flavor-violating Yukawa interactions and the mass spectrum of the additional Higgs bosons.

\begin{itemize}
    \item {\bf Same-sign di-top final state: } Same-sign top-quark production is one of the most promising signatures of the scenario considered here. The signal predominantly arises from the associated production of $H_2$ or $H_3$ with a top quark, followed by the flavor-violating decay of the neutral Higgs boson:
\begin{equation}
c g \to t\, H_i \to t(t \bar{c}), \qquad i=2,3,
\label{eq}
\end{equation}
The charge-conjugate process gives rise to the corresponding anti-top final state. When both same-sign top quarks decay leptonically, these processes produce same-sign dilepton (SSDL) signatures accompanied by $b$-tagged jets and additional hadronic activity. Owing to their relatively small SM backgrounds, such signatures provide sensitive probes of flavor-changing neutral Higgs interactions.
Similar processes have been studied in the general 2HDM~\cite{Kohda:2017fkn,Hou:2018zmg,Hou:2024bzh,Hou:2025tjp}. In the general 2HDM, the flavor-violating Yukawa couplings are generally treated as independent parameters. In contrast, in the SCPV 2HDM considered here, these couplings are correlated through the underlying Yukawa structure and the requirement that CP violation originates spontaneously.

\item {\bf Three-top final state:} For sufficiently heavy $H_3$, the decay $H_3\to t\bar t$ leads to three-top production:
\begin{equation}
cg \to t H_3 \to t (t \bar t). 
\end{equation}
The subsequent top-quark decays produce final states containing multiple $b$-tagged jets and leptons. Although the branching ratio of $H_3\to t\bar t$ is typically smaller than that of the dominant flavor-violating or bosonic decay channels, three-top production provides a distinctive signature with complementary sensitivity to the Yukawa interactions. It can also contribute to SSDL event samples.

\item {\bf Cascade channels:} The presence of multiple additional Higgs bosons allows for bosonic cascade decays such as
\begin{equation}
H_3 \to H_2 Z, \qquad H_3 \to H^\pm W^\mp,
\label{eq}
\end{equation}
When combined with associated top-quark production, these decays give rise to processes such as
\begin{equation}
c g \to tH_3 \to t(H_2 Z) \to t(tc)Z, \qquad
c g \to tH_3 \to t(H^\pm W^\mp).
\label{eq}
\end{equation}
Similar cascade signatures can also arise from gluon-fusion production of $H_3$, followed by its decay into $H_2Z$ or $H^\pm W^\mp$.
Although these channels may have smaller rates than the dominant fermionic modes, their final states contain additional electroweak gauge bosons and can provide valuable information on the scalar mass hierarchy. They therefore offer complementary probes of the extended Higgs sector.

\end{itemize}

\begin{center}
\begin{figure}[t]
\includegraphics[width=7cm]{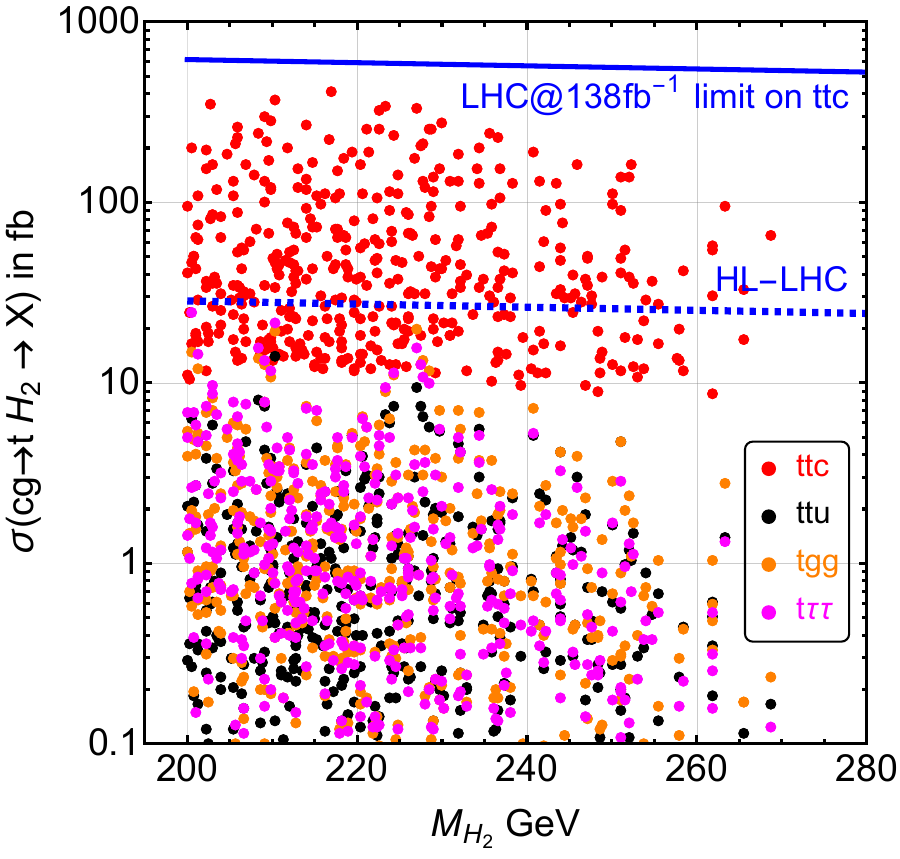}~~
\includegraphics[width=7cm]{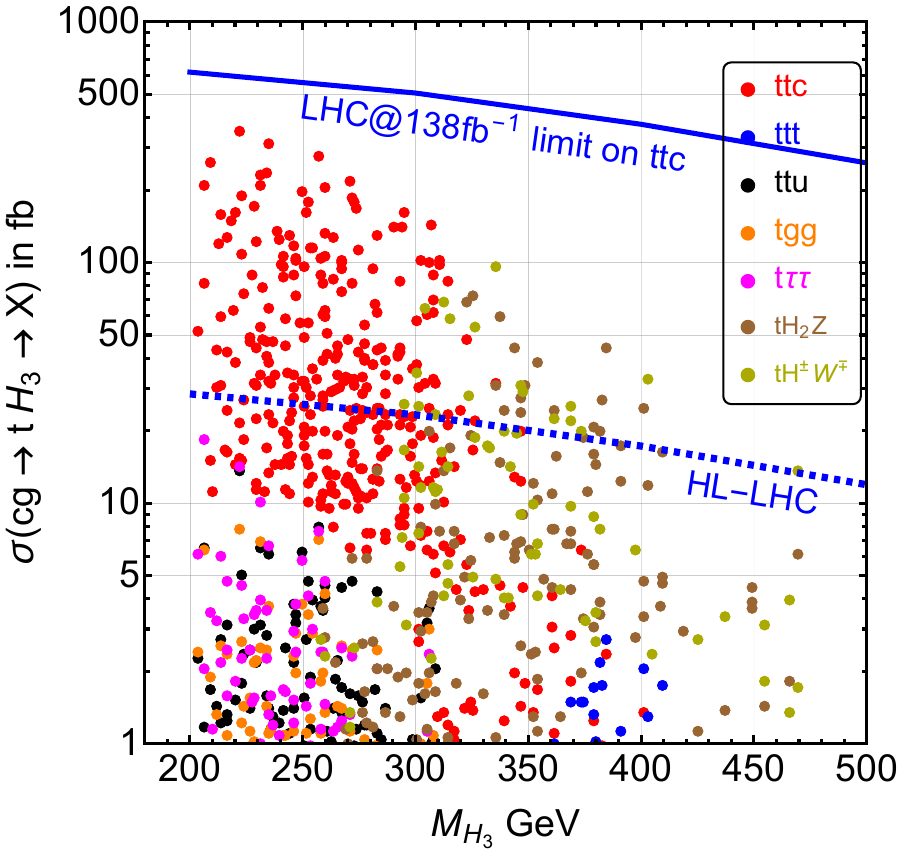}
\caption{Predicted cross sections for $pp\to tH_i+\bar tH_i$ ($i=2,3$), followed by the decays of $H_i$, at the 13 TeV LHC. The left and right panels correspond to $H_2$ and $H_3$, respectively. All parameter points satisfy the theoretical and experimental constraints. The solid blue curves indicate the current upper limits on the same-sign top-quark signal, while the dashed blue curves show the projected HL-LHC sensitivity for an integrated luminosity of $3000\,\mathrm{fb}^{-1}$.} 
\label{fig:h2h3-pheno}
\end{figure}
\end{center}

To estimate the production cross sections of the characteristic collider signatures, we implement the SCPV 2HDM in \texttt{FeynRules-2.3}~\cite{Alloul:2013bka} and generate the corresponding model files. The leading-order production cross sections are calculated using \texttt{MadGraph5-aMC@NLO-3.5.3}~\cite{Alwall:2011uj,Alwall:2014hca}.

Fig.~\ref{fig:h2h3-pheno} shows the predicted cross sections for the associated production of $H_2$ (left panel) and $H_3$ (right panel) with a top quark, followed by their subsequent decays, at the 13 TeV LHC. The results correspond to parameter points satisfying all the theoretical and experimental constraints discussed in Sec.~\ref{sec:constraints}.
The solid blue curves indicate the current upper limits on the same-sign top-quark signal $tt\bar c+\bar t\bar t c$, obtained by translating the experimental bounds on $\rho_{tc}$ from the LHC searches for SSDL events~\cite{ATLAS:2023tlp,CMS:2023xpx} into the corresponding signal cross sections.
The dashed blue curves represent the projected sensitivity at the HL-LHC with an integrated luminosity of $3000\,\mathrm{fb}^{-1}$, obtained by extrapolating the current constraints.
A substantial fraction of the parameter points that remain viable under the current constraints can be probed at the HL-LHC through searches for same-sign top-quark production. This highlights the importance of the flavor-violating top-charm Yukawa interaction as a characteristic collider signature of the SCPV 2HDM.

\begin{center}
\begin{figure}[t]
\includegraphics[width=7cm]{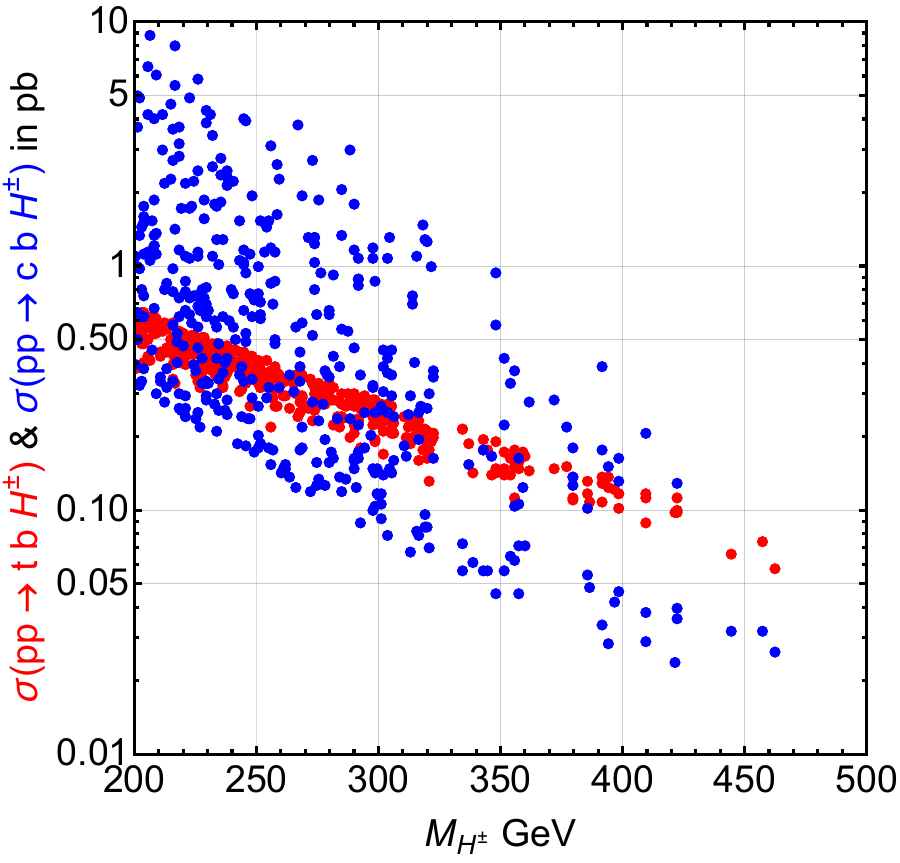}~~
\includegraphics[width=7cm]{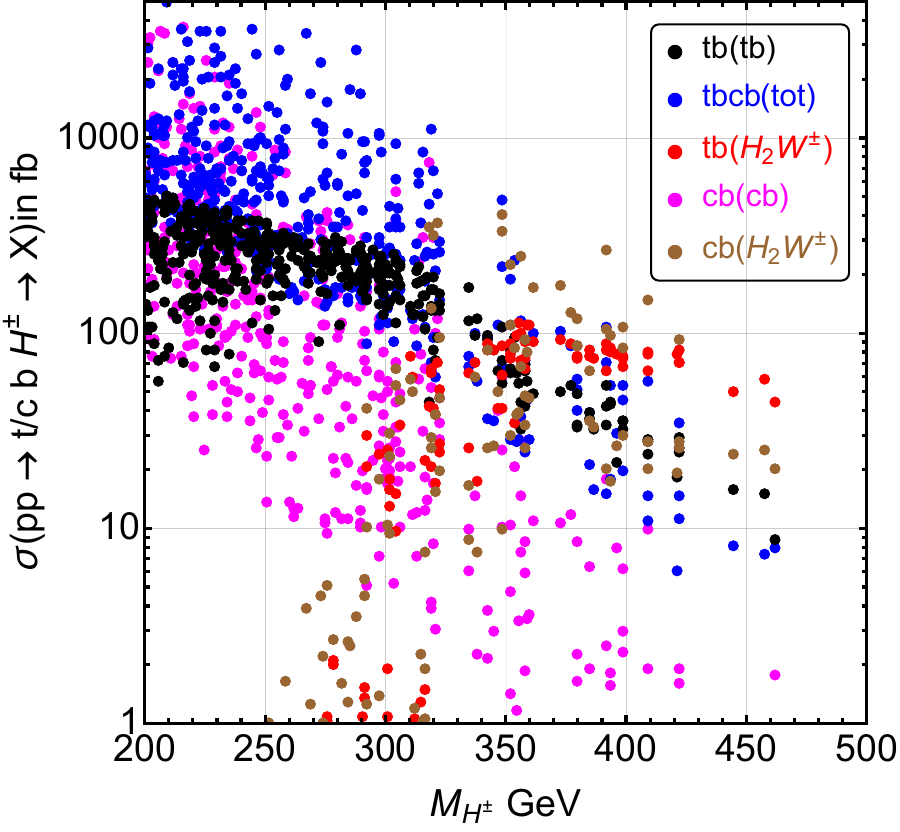}
\caption{Production cross sections of charged Higgs bosons at the $13~\mathrm{TeV}$ LHC. The left panel shows the cross sections for $pp\to tbH^\pm$ and $pp\to cbH^\pm$, while the right panel shows the cross sections for various final states arising from charged Higgs decays. The results are shown as functions of the charged Higgs mass for parameter points satisfying all theoretical and experimental constraints. The final states indicated in parentheses correspond to the decay products of $H^\pm$.} 
\label{fig:hpm-pheno}
\end{figure}
\end{center}

\subsection{Signatures of the Charged Higgs Bosons}

We next investigate the collider signatures associated with charged Higgs boson production. As discussed above, the relevant production channels are $pp\to tbH^\pm$ and $pp\to cbH^\pm$, followed by the fermionic decays $H^\pm\to tb,cb$ or the bosonic decay $H^\pm\to H_2W^\pm$.

The left panel of Fig.~\ref{fig:hpm-pheno} shows the production cross sections for $pp\to tbH^\pm$ and $pp\to cbH^\pm$ as functions of the charged Higgs mass. The $cbH^\pm$ production mode tends to dominate over a substantial part of the viable parameter space, reflecting the characteristic flavor structure of the Yukawa interactions. The relative importance of the two production modes depends on the Yukawa couplings and the available phase space.
Combining these production and decay channels, we obtain several characteristic final states:
\begin{equation}
tb\,H^\pm \to tb\,(tb\,/\,cb\,/\,H_2W^\pm), \qquad
cb\,H^\pm \to cb\,(tb\,/\,cb\,/\,H_2W^\pm). 
\label{eq}
\end{equation}
The right panel of Fig.~\ref{fig:hpm-pheno} shows the corresponding cross sections for the various final states. At relatively low charged Higgs masses, the fermionic decay channels play an important role. As the charged Higgs mass increases, the bosonic decay $H^\pm\to H_2W^\pm$ becomes accessible and can provide a significant contribution to the collider signatures.
These charged Higgs channels offer complementary probes of the Yukawa structure and scalar mass spectrum. In particular, final states involving $H_2W^\pm$ can produce multiple top quarks and electroweak gauge bosons through the subsequent decay of $H_2$. Related charged Higgs signatures in the general 2HDM have been investigated in Refs.~\cite{Ghosh:2019exx,Hou:2021xiq,Hou:2024ibt,Fang:2025nma}.

\begin{center}
\begin{figure}[t]
\includegraphics[width=7.4cm]{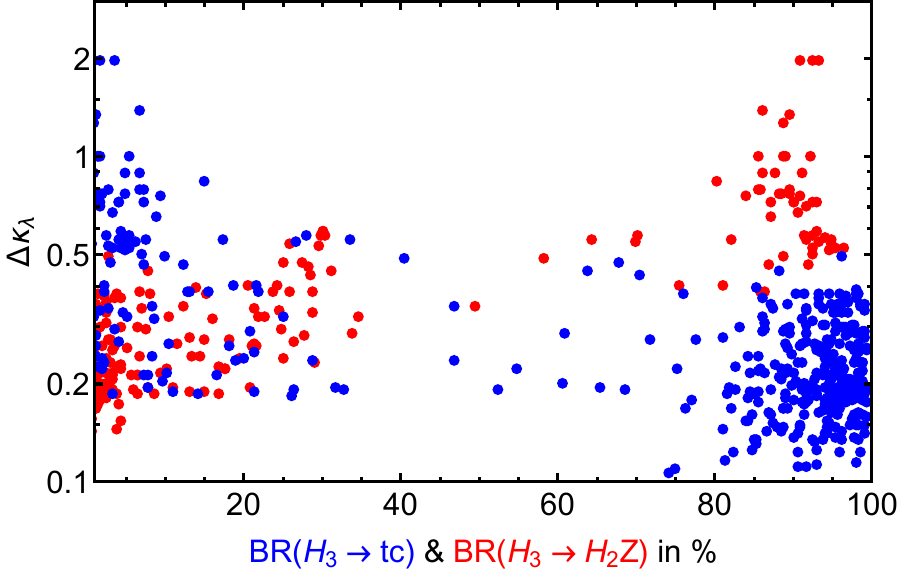}~~
\includegraphics[width=7.4cm]{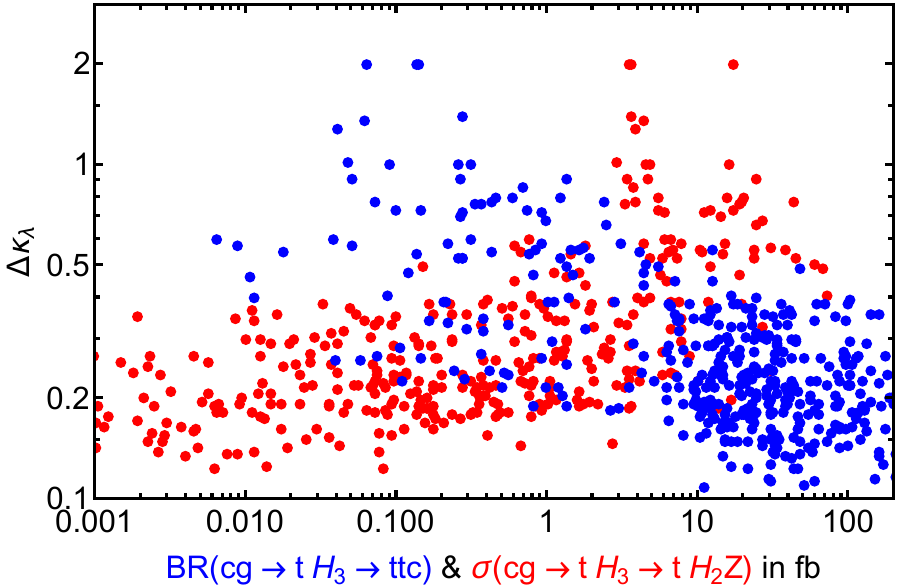}
\caption{Correlations between the relative deviation in the SM-like Higgs trilinear self-coupling, $\Delta\kappa_h$, and collider observables associated with $H_3$. The left panel shows the correlation with the branching ratios of $H_3\to tc$ and $H_3\to H_2Z$, while the right panel shows the correlation with the cross sections for the $ttc$ and $tH_2Z$ final states originating from associated $tH_3$ production. All parameter points satisfy the theoretical and experimental constraints.} 
\label{fig:corr}
\end{figure}
\end{center}

\subsection{Correlation with SM Higgs Measurement}

The collider phenomenology of the additional Higgs bosons is closely related to precision measurements of the SM-like Higgs boson self-coupling, $\lambda_{hhh}$. As discussed in Sec.~\ref{sec:constraints}, the additional Higgs bosons induce non-decoupling loop corrections to $\lambda_{hhh}$, which can lead to sizable deviations from the SM prediction even in the Higgs alignment limit.
Such deviations may be probed through measurements of Higgs boson pair production at the HL-LHC~\cite{Cepeda:2019klc} and future lepton colliders, including the ILC~\cite{ILC:2019gyn} and muon colliders~\cite{Han:2020pif}.
Since the non-decoupling corrections depend strongly on the masses of the additional Higgs bosons, they are naturally correlated with their decay patterns and production rates. In particular, the mass hierarchy among $H_2$, $H_3$, and $H^\pm$ affects both the magnitude of the Higgs self-coupling deviation and the branching ratios of the bosonic cascade decays of $H_3$.

Fig.~\ref{fig:corr} illustrates these correlations in terms of the relative deviation $\Delta\kappa_h$, defined as
\begin{align}
\Delta\kappa_h \equiv \frac{\lambda_{hhh}^{\rm 1-loop}-\lambda_{hhh}^{\rm 1-loop}\Big|_{\rm SM}}{\lambda_{hhh}^{\rm 1-loop}\Big|_{\rm SM}}. 
\end{align}
The left panel shows the correlation between $\Delta\kappa_h$ and the branching ratios of $H_3\to t\bar c+\bar t c$ and $H_3\to H_2Z$. The right panel shows the corresponding correlation with the cross sections for $cg\to tH_3\to tt\bar c$ and $cg\to tH_3\to tH_2Z$, together with their charge-conjugate processes.
As discussed in Ref.~\cite{Mondal:2025iie}, sizable deviations in the Higgs self-coupling can arise in regions with a hierarchical additional Higgs mass spectrum. Such mass hierarchies also modify the relative importance of the fermionic and bosonic decay channels of $H_3$, leading to characteristic correlations between $\Delta\kappa_h$ and the collider observables shown in Fig.~\ref{fig:corr}.

These results demonstrate that direct searches for additional Higgs bosons and precision measurements of the SM-like Higgs self-coupling provide complementary probes of the SCPV 2HDM. A combined analysis of these observables can therefore help test the characteristic non-decoupling structure of the extended Higgs sector.

In summary, the SCPV 2HDM predicts characteristic collider signatures arising from flavor-violating Yukawa interactions and bosonic cascade decays of the additional Higgs bosons. Same-sign top-quark production provides a promising probe of the top-charm Yukawa interaction, while cascade decays involving electroweak gauge bosons offer complementary information on the additional Higgs mass spectrum. Furthermore, these signatures are correlated with non-decoupling corrections to the SM-like Higgs self-coupling. The combined exploration of these observables at the LHC and future colliders therefore provides a powerful means of testing the SCPV scenario.


\section{Discussion and Conclusion \label{sec:discussions}}

We have investigated the phenomenology of spontaneous CP violation (SCPV) in the general two-Higgs-doublet model (2HDM), with particular emphasis on the interplay between flavor-violating interactions of the additional Higgs bosons and precision observables of the discovered Higgs boson. In this framework, all parameters in the Higgs potential and Yukawa interactions are real, and CP violation arises solely from the relative phase between the vacuum expectation values of the two Higgs doublets. The stationary conditions of the Higgs potential constrain its dimensionful parameters to be of the electroweak scale, thereby preventing the additional Higgs bosons from decoupling. Moreover, reproducing the observed Kobayashi–Maskawa (KM) phase in the Cabibbo–Kobayashi–Maskawa (CKM) matrix requires flavor misalignment in at least one of the quark sectors. These features lead to a highly constrained realization of the general 2HDM, in which the scalar and Yukawa sectors are linked by the spontaneous origin of CP violation.

We have considered a scenario in which the down-type quark and charged-lepton Yukawa matrices are flavor aligned, while flavor-violating interactions are allowed in the up-type quark sector. We have also imposed the Higgs alignment limit, where the tree-level couplings of the discovered Higgs boson coincide with their SM predictions. We have explored the viable parameter space by imposing theoretical constraints from vacuum stability and perturbative unitarity, together with experimental constraints from electroweak precision observables, flavor measurements, electric dipole moments (EDMs), Higgs signal strengths, and direct searches for additional Higgs bosons. In particular, the electron EDM imposes a stringent constraint, which can be satisfied through destructive interference between the scalar- and fermion-loop Barr–Zee contributions.
We have found that sizable flavor-violating top–charm Yukawa couplings can survive these constraints, together with a sizable diagonal top Yukawa coupling. These interactions give rise to characteristic collider signatures of the additional Higgs bosons. In particular, the associated production processes $cg\to tH_{2,3}$ lead to same-sign top-quark and three-top final states, while the cascade decays $H_3\to H_2Z$ and $H_3\to H^\pm W^\mp$ provide complementary signatures sensitive to the additional Higgs mass spectrum. Furthermore, the flavor-violating Yukawa structure can enhance charged Higgs production in association with charm and bottom quarks, offering another promising avenue for probing the scenario at the LHC.

A distinctive consequence of SCPV is the non-decoupling nature of the additional Higgs sector, which leads to sizable deviations in precision Higgs observables even in the Higgs alignment limit. In particular, charged Higgs loop contributions to the $h\to\gamma\gamma$ and $h\to Z\gamma$ decay rates remain finite in the large charged Higgs mass limit, resulting in asymptotic deviations of approximately $-10\%$ and $-4\%$, respectively, from the SM predictions. The non-decoupling effects also induce potentially large one-loop corrections to the trilinear Higgs self-coupling $\lambda_{hhh}$, whose leading contributions exhibit a quartic dependence on the additional Higgs boson masses. We have further demonstrated correlations between the deviations in $\lambda_{hhh}$ and the decay patterns and production signatures of the additional neutral Higgs bosons. These correlations provide a direct connection between precision measurements of the discovered Higgs boson and searches for the additional Higgs states.

Our results highlight the importance of combining direct searches for flavor-violating and cascade-decay signatures of additional Higgs bosons with precision measurements of the discovered Higgs boson. Together with complementary constraints from flavor observables and EDMs, such measurements provide a powerful means of probing the correlated scalar and Yukawa structures predicted by SCPV in the 2HDM.

\section*{Acknowledgements}
T.M. was supported by the Anusandhan National Research Foundation (ANRF) (Grant Number: ANRF/ECRG/2025/002387/PMS).

\bibliography{references2}

\end{document}